\documentclass[useAMS,usenatbib]{mn2e}

\usepackage{hyperref}
\usepackage{hyperref,xcolor}

\usepackage{graphicx}
\usepackage{amssymb}
\usepackage{amsmath}
\usepackage[normalem]{ulem}

\definecolor{winered}{rgb}{0.5,0,0}
\definecolor{mousegrey}{rgb}{0.2,0.2,0.7}
\definecolor{violet}{rgb}{0.2,0.2,0.7}

\hypersetup{pdftitle={},
  pdfauthor={Ben Granett},
  hyperfootnotes=true,
  colorlinks=true,
  pdfborder={0 0 0},
  citecolor=black,
  urlcolor=blue,
  filecolor=blue,
  linkcolor=black,
}

\newcommand{\muK}{$\mathrm{\mu K}$}

\newcommand{\tcmb}{T_{\small \rm CMB}}

\newcommand{\xbar}{\overline{x}}

\newcommand{\simlt}{\,\hbox{\lower0.6ex\hbox{$\sim$}\llap{\raise0.6ex\hbox{$<$}}}\,}

\newcommand{\LCDM}{$\Lambda$CDM}

\newcommand{\hmpc}{h{\rm Mpc^{-1}}}
\newcommand{\mpch}{h^{-1}{\rm Mpc}}

\newcommand{\nside}{n_{\rm side}}

\defcitealias{Granett08}{GraNeSz}

\title[BOSS ISW]{The Integrated Sachs-Wolfe Signal from BOSS Super-Structures}

 \author[B.R.~Granett, A. Kov\'acs, A.J.~Hawken]{Benjamin R. Granett$^{1}$\thanks{E-mail:\href{mailto:ben.granett@brera.inaf.it}{ben.granett@brera.inaf.it}},
  Andr\'as Kov\'acs$^{2}$, 
  Adam J. Hawken$^{1}$    \\
   $^{1}$Istituto Nazionale di Astrofisica - Osservatorio Astronomico di Brera, Via E. Bianchi 46, 23807 Merate, Italy\\ 
   $^2$ Institut de F\'isica d'Altes Energies, Universitat Aut\'onoma de Barcelona, E-08193 Bellaterra (Barcelona), Spain
}

\begin{document}


\pagerange{\pageref{firstpage}--\pageref{lastpage}} \pubyear{2011}

\maketitle
\label{firstpage}

\begin{abstract}
Cosmic structures leave an imprint on the microwave background radiation through the integrated Sachs-Wolfe effect.  We construct a template map of the linear signal using the SDSS-III Baryon Acoustic Oscillation Survey at redshift $0.43 < z < 0.65$.  We verify  the imprint of this map on the Planck CMB temperature map at the 97\% confidence level and show consistency with the density-temperature cross-correlation measurement.  Using this ISW reconstruction as a template we investigate the presence of ISW sources and further examine the properties of the Granett-Neyrinck-Szapudi supervoid and supercluster catalogue.  
We characterise the three-dimensional density profiles of these structures for the first time and demonstrate that they are significant structures.    Model fits demonstrate that the supervoids are elongated along the line-of-sight and we suggest that this special orientation may be picked out by the void-finding algorithm in photometric redshift space.   
We measure the mean temperature profiles in Planck maps from public void and cluster catalogues.
  In an attempt to maximise the stacked ISW signal we construct a new catalogue of super-structures based upon local peaks and troughs of the gravitational potential.  However, we do not find a significant correlation between these structures and the CMB temperature. 

\end{abstract}
\begin{keywords}
cosmological parameters, observations, large-scale structure of
Universe, methods: statistical
\end{keywords}

\section{Introduction}

Striking correlations have been observed between cosmic super-structures and the cosmic microwave background (CMB).  Most prominently \citet{Granett08} (hereafter \citetalias{Granett08}) reported a $4\sigma$ detection of voids and clusters at redshift $z=0.5$ identified in the Sloan Digital Sky Survey (SDSS)   imprinted on the CMB as 10\muK{} cold and hot spots on 4\degr{} scales.  Additionally, there is evidence for a supervoid at redshift $z=0.2$ aligned with the Cold Spot suggesting a causal relationship \citep{Szapudi15}.    While these findings may be statistically significant, the physical mechanism producing the correlation has yet to be determined.
The leading candidate has been the late-time integrated Sachs-Wolfe effect \citep{sachswolfe,reessciama}.  However, the scale of the temperature correlation is inconsistent with the expectation in the standard \LCDM{} cosmological model for typical supervoid and superclusters \citep{Cai10,Nadathur12,HM13,Flender13,Nadathur14,Hotchkiss15,Aiola15} and cannot fully explain the Cold Spot \citep{Nadathur14cs,Zibin14}.   Moreover, the  signal-to-noise ratio for the  ISW measurement with SDSS galaxies  is expected to be $\sim1.7$ \citep{HM14}, while a full-sky survey to redshift 2 can reach a maximum  $S/N\sim7$ \citep{Afshordi04}.   Thus the anomalous  contribution from individual sources may point to new physics. 

The ISW signal is challenging to measure because the amplitude is an order of magnitude below the primary temperature anisotropies.  It may be detected by 
cross-correlating galaxy tracers with the cosmic microwave background over the sky or at the locations of particular super-structures (\citet{Nishizawa14} offers a review).    In flat cosmologies the ISW signal is sensitive to dark energy \citep{Crittenden96} and provides important constraints for modified gravity models \citep{Giannantonio10} as well as for primordial non-Gaussianity \citep{Afshordi08}.  The signal may be enhanced by adding constraints from the primordial polarisation of the CMB \citep{Frommert09}.  It may also be independently estimated by correlating against the infrared background \citep{Ilic11} and 21cm brightness maps \citep{Raccanelli15} adding to the detection significance.

Since the ISW effect is subtle, analyses can be prone to over-fitting and suffer from the `look elsewhere' effect \citep{Peiris2014}.  It is a particular problem for the study of ISW sources since the physical nature of supervoid and supercluster structures identified in galaxy surveys is not well-known and there is not a clear expectation for the signal.
  Numerical N-body simulations have been used to predict the ISW anisotropy arising from voids and clusters and to determine the optimal detection filters \citep{Cai10,Cai14,Hotchkiss15}.  Guided by these results, attempts to reproduce the \citetalias{Granett08} measurement with other super-structure catalogues have lead to only marginal or null detections  \citep{Kovacs15,Hotchkiss15,Nadathur14,Cai14,Ilic13}.   These studies would suggest that the \citetalias{Granett08} detection arises from a different physical mechanism or is a statistical fluke.

However, an important piece of the puzzle remains out of place.  To date, there has been no characterisation of the three-dimensional properties of the \citetalias{Granett08} structures.  Analyses have used superstructure catalogues derived from spectroscopic redshift catalogues employing variations of the watershed algorithm based on the ZOBOV code \citep{Neyrinck08,Neyrinck05}.   The \citetalias{Granett08} super-structures were identified with a photometric redshift catalogue and the correspondence between the void finding techniques is not yet clear.   As pointed out by \citet{Kovacs15}  the \citetalias{Granett08} supervoids  do not correlate with the voids identified in three-dimensions with the watershed algorithm.  

Besides the use of N-body simulations, an alternative  empirical approach is possible: we may predict the ISW signal directly from galaxy surveys, without modelling
individual structures.  All-sky reconstructions of the ISW signal assuming linear dynamics have already proven to be useful for investigating possible large-scale anomalies on the CMB  \citep{Maturi07,Granett09,Francis10,Rassat13,Manzotti14}.

Previously, the \citetalias{Granett08} sources were investigated using an ISW map constructed based upon the photometric luminous red galaxy (LRG) sample at redshift $z=0.5$ \citep{Granett09}.  Although the map contained hot and cold spots corresponding to the super-structures, the amplitude of the effect was slight, with only $\Delta T^{cluster} - \Delta T^{void}=0.08\pm0.1$\muK{}.
This result seemingly contradicted the claim that the 10\muK{} anisotropy was arising from the linear ISW effect.  To resolve this question we are motivated to improve the fidelity of the ISW reconstruction by building it using a galaxy sample with measured spectroscopic redshifts. 

We are now in a position to carry out these investigations using the SDSS-III Baryon Acoustic Oscillation Survey CMASS spectroscopic sample which covers the sky area and redshift range of the original \citetalias{Granett08} analysis \citep{DR12,Bolton12}.  In general, the BOSS sample is interesting for ISW studies because, due to the volume dependence, the sensitivity to ISW is expected to peak at  redshift $z\sim0.5$ \citep{Afshordi04}.  Analyses have shown the cross-correlation measurements of the luminous red galaxy sample in the North Galactic cap show an excess signal \citep{Ho08,Giannantonio08,Granett09}, despite low significance \citep{Sawangwit10}. The complete photometric BOSS sample was analysed by \citet{HM14} and \citet{Giannantonio14}.  

Galaxies in the CMASS sample have spectroscopic redshift measurements, as opposed to photometric redshifts in the photometric LRG samples used previously.  While precise redshift measurements are not necessary for  cross-correlation measurements using the projected galaxy density field,  they are necessary to reconstruct the three-dimensional potential.

We begin this article by detailing the spectroscopic CMASS sample in Sec. \ref{sec:data} including our corrections for the known systematic inhomogeneities.  In Sec. \ref{sec:iswmap} we then describe the  procedure to estimate the underlying density field, the three-dimensional gravitational potential and the linear ISW map.  We validate the ISW map using the template 
fit statistic against the Planck CMB temperature map.  In Sec. \ref{sec:crosscorr} we carry out the consistency check between the ISW map and the angular galaxy-CMB cross correlation function.  In Sec. \ref{sec:radprof} we characterise the \citetalias{Granett08} super-structures by fitting the three-dimensional density profiles.  In Sec. \ref{sec:super} we address our main concerns, ISW sources.  We characterise the three-dimensional properties of the \citetalias{Granett08} structures using the CMASS sample and estimate their contribution to the linear ISW effect using the ISW map.  For comparison, we  consider the publicly available void catalogue by \citet{Sutter14}.  We also construct a new super-structure catalogue based upon the local peaks and troughs of the three-dimensional gravitational potential that should be ideal ISW sources.  We conclude in Sec. \ref{sec:concl} with suggestions for future investigations.

\begin{figure}
\begin{center}
\includegraphics[scale=0.9]{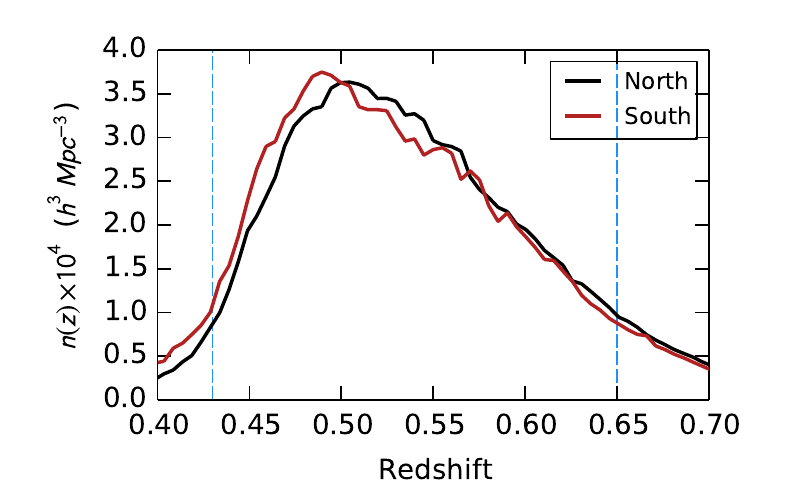}
\caption{Redshift distribution of the CMASS samples In the North and South Galactic cap fields.  The vertical dashed lines indicate the redshift ranges used in this study. \label{fig:nz}}
\end{center}
\end{figure}

\begin{figure}
\begin{center}
\includegraphics[scale=.9]{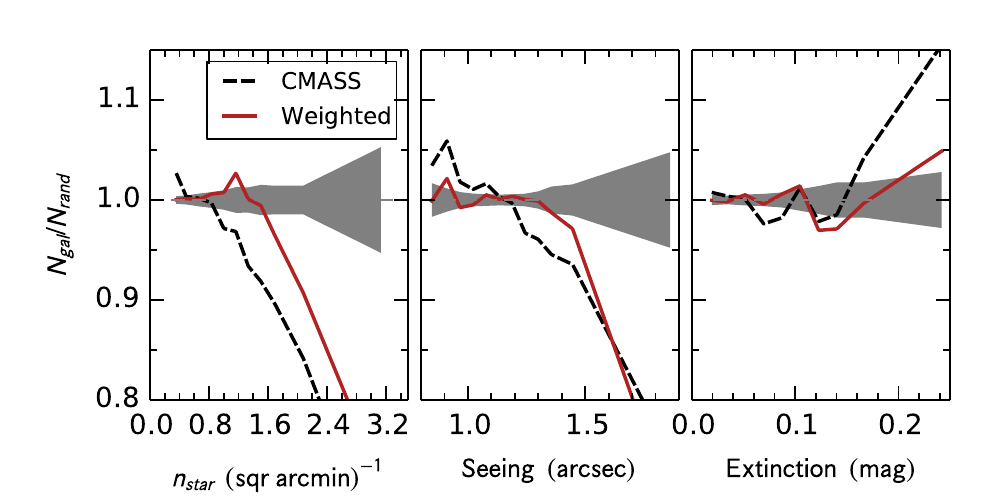}
\caption{The mean number density of the CMASS sample with respect to a uniform random catalogue as a function of stellar density (left), seeing (center) and Galactic extinction (right). The dashed line shows the systematic trends present in the catalogue. After applying corrective weights we find the trends indicated by the solid lines. The shaded regions indicate the 68\% uncertainty range for Poisson counts. \label{fig:sys}}
\end{center}
\end{figure}

\begin{figure}
\begin{center}
\includegraphics[scale=0.5]{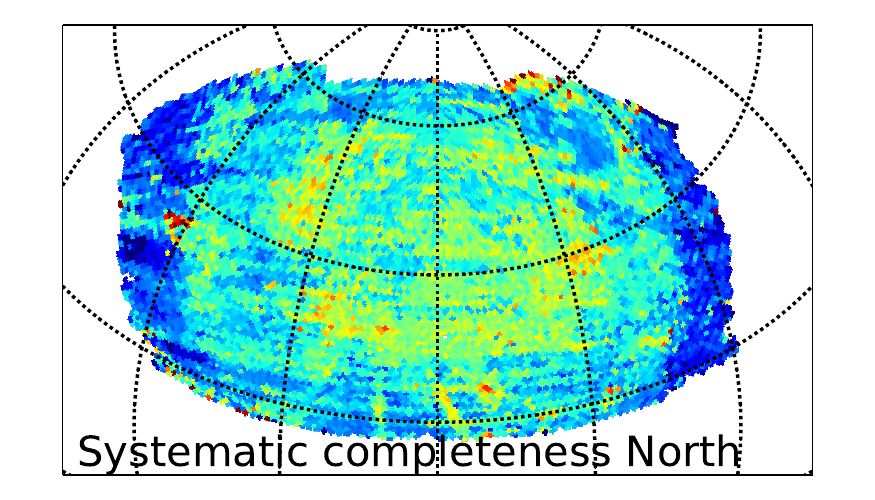}\hspace*{-0.4cm}\includegraphics[scale=0.5]{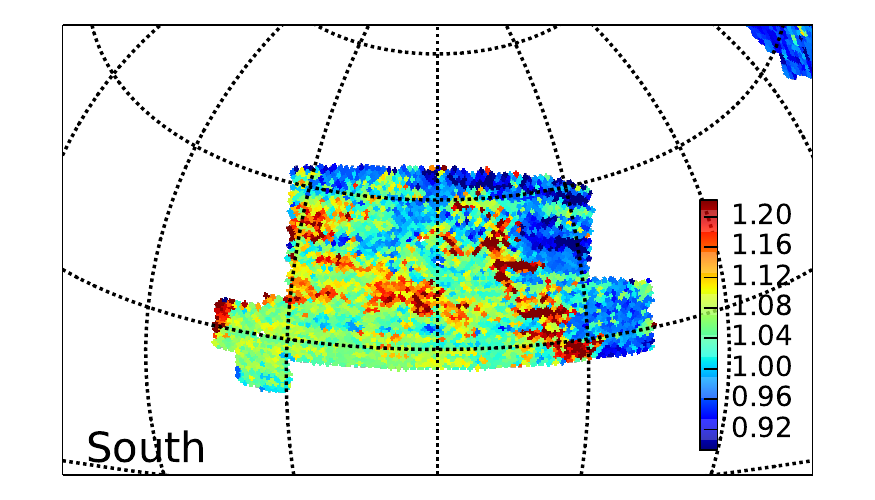}
\includegraphics[scale=0.5]{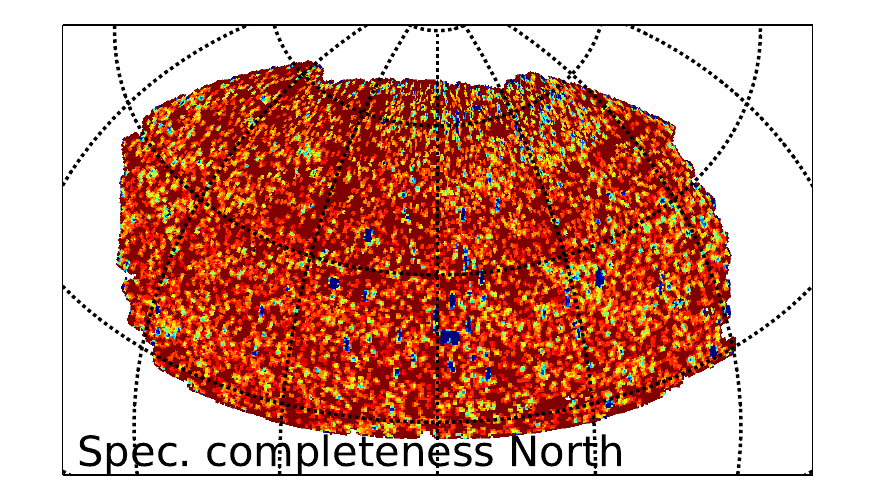}\hspace*{-0.4cm}\includegraphics[scale=0.5]{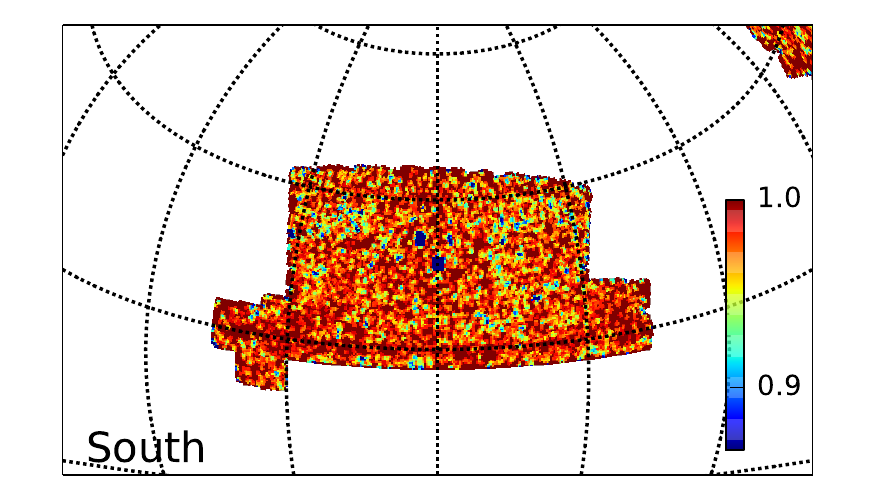}
\caption{The CMASS  completeness maps for the North and South galactic caps.  The top frames show the systematic completeness correlated with stellar density, seeing, extinction and galaxy surface brightness.  The bottom frames show the spectroscopic completeness. \label{fig:tsr}}
\end{center}
\end{figure}

\section{Data}\label{sec:data}
\subsection{Galaxy sample}
We use the constant stellar mass (CMASS) galaxy sample from the Sloan
Digital Sky Survey-III (SDSS-III) Baryon Oscillation Spectroscopic
Survey (BOSS) Data Release 12 \citep{DR12,Bolton12}.    We extract the
photometric parent catalog from the targeting catalog for the ninth and final run.
This provides a uniform photometric selection over
the full survey.  The targets are selected
with the CMASS target selection flag and fibre magnitude limit ({\tt
  boss\_target1\&BOSS\_CMASS>0}).  We then look up the redshifts of
these sources in the spectroscopic objects database selecting all
primary objects ({\tt scienceprimary=1}).  We match the photometric
and spectroscopic catalogs using position on the sky with a match
radius of 1 arcsec.  We take the BOSS galaxy redshift estimate {\tt
  z\_noqso} when successfully measured ({\tt zwarning\_noqso=0}).  If
the BOSS redshift estimate is not available, we take the legacy
redshift measurement as long as it was successfully measured ({\tt
  zwarning=0}).

The CMASS sample covers two discontiguous fields in the Northern and Southern Galactic caps.
We use the survey mask that characterises the CMASS targeting
catalogue. We extract the polygons from the mask that correspond to
the BOSS survey runs (2,5,7,8 and 9) based upon the chunk identifiers.
We exclude sources that fall within the photometric quality,
bright star and centre post masks.  We use the Mangle code
\citep{Mangle} to carry out these tasks.

The spectroscopic redshift distribution of the final catalog is shown
in Fig. \ref{fig:nz}.  When estimating the underlying density field we combine
 two subsamples with redshift ranges (I) $0.43<z<0.55$ and (II) $0.55<z<0.65$ described in Table.
\ref{tab:samples}.  We consider separately the Northern and Southern Galactic cap fields.

\subsection{Survey completeness}
The parent catalog from which CMASS targets were selected is known to
be inhomogeneous.  Seeing conditions, stellar density and Galactic
extinction were found to impact the number density of galaxies with a
dependence on galaxy surface brightness \citep{Ho12,Ross12}.  Here, we
quantify those effects and compute corrective weights.

We first estimate the spectroscopic sampling rate.  The number of
sources in the photometric target sample is given by $N_{targ}$ of
which $N_{obs}$ have measured redshifts.  
After classifying as stars or galaxies we have
$N_{obs}=N_{gal}+N_{star}$.  We then define the spectroscopic completeness
for galaxies as $c_{spec}=N_{gal}/ (N_{targ}-N_{star})$.  The
completeness is computed for each galaxy by counting the number of
sources within a radius of $R=30$ arcmin.  From the completeness we
derive a weight for each galaxy: $w_{spec,i}=1/c_{spec,i}$.  The map
of the spectroscopic completeness is shown in Fig. \ref{fig:tsr}.
At redshift $z=0.5$, $30$ arcmin on the sky corresponds to a comoving separation of $11 \mpch$, so we do not expect effects correlated on the scale of the detector to influence large-scale modes in the density field.

To address possible large-scale systematic effects, we construct maps of each
potential source.  We selected stars as in \citet{Anderson14} by
taking sources with $17.5<i_{\rm dered}<19.9$ from the SDSS {\tt star}
database table to create an angular density healpix map \citep{Healpix}.  We build a
map of the seeing taking the $i$-band FWHM of the PSF from the target
photometric catalog averaging the seeing values within Healpix
cells.  We create a galactic extinction map in the same way using the
$i$ band extinction.  We also use the 2" fibre magnitude as a proxy
for galaxy surface brightness.

We compare the density of galaxies to the density of unclustered random points
distributed  over the survey area.  Each random point has an
associated stellar density, seeing and extinction assigned based upon
the sky coordinate and a surface brightness drawn independently from
the observed distribution.  Fig. \ref{fig:sys} shows the normalised
galaxy density with respect to randoms as a function of each
potential systematic.  There is a deficit of sources in regions with
high stellar density or poor seeing while we find an excess of sources in regions with high values
of reddening.

The completeness depends on a four-dimensional parameter space of
fibre magnitude, stellar density, seeing and extinction.  
We develop a non-parametric approach using the $K$-means algorithm to
adaptively bin the data.  We use the algorithm to construct $K=500$
adaptive bins each associated with roughly 2000 unique galaxies.  This is
carried out using a rank-order distance metric to standardise the
parameters.

We compute the completeness in each bin by counting the weighted
number of galaxies with respect to the number of randoms:
$c_{sys}=\sum_{i=1}^N w_{spec,i} \alpha / N_{r}$.  The completeness is
then assigned to each galaxy in the bin.  The map of the systematic
completeness is shown in the top row of Fig. \ref{fig:tsr}.

After weighting the density field by the inverse of the systematic
completeness, we find the trends given by the solid red curve in
Fig. \ref{fig:sys}.  The correlations are reduced although we still
see a significant dependence at the parameter extremes.  However, the
last bin in the plots is an average over 150 cells, so it represents only
1\% of the total area.  
The trends shown in Fig. \ref{fig:sys} are in good agreement with those published in \citet{Ross12}  
which gives us confidence in our independent methodology.
The total selection function is given by the
product of the spectroscopic and systematic completeness fractions:
$c_i = c_{spec,i}\times c_{sys,i}$.

\begin{figure*}
\begin{center}
\includegraphics[scale=1]{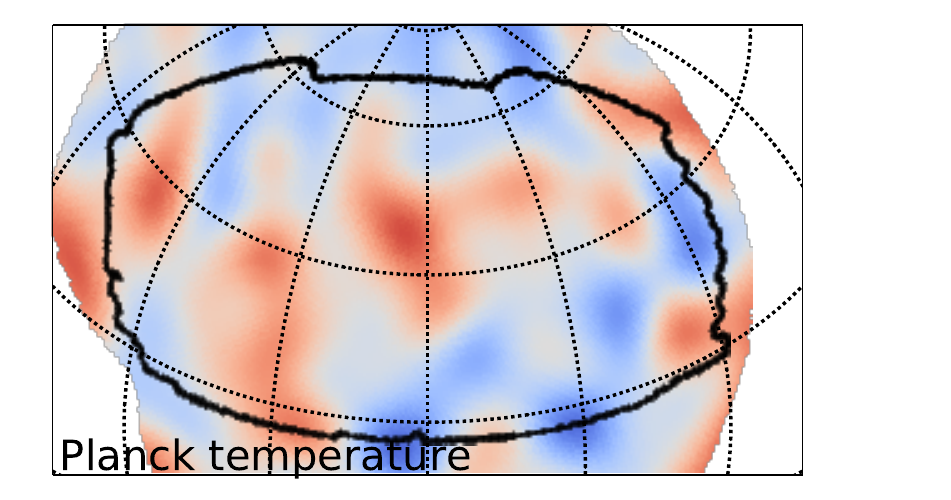}\hspace*{-1.9cm}\includegraphics[scale=1]{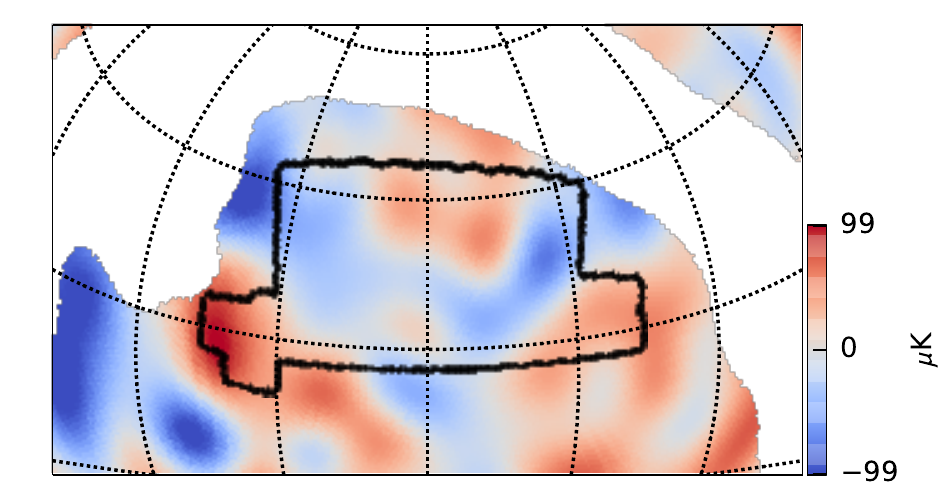}
\includegraphics[scale=1]{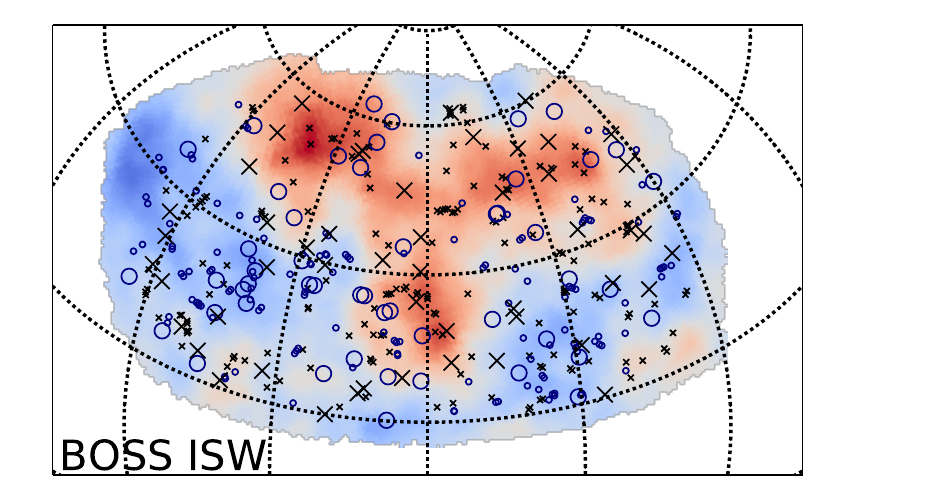}\hspace*{-1.9cm}\includegraphics[scale=1]{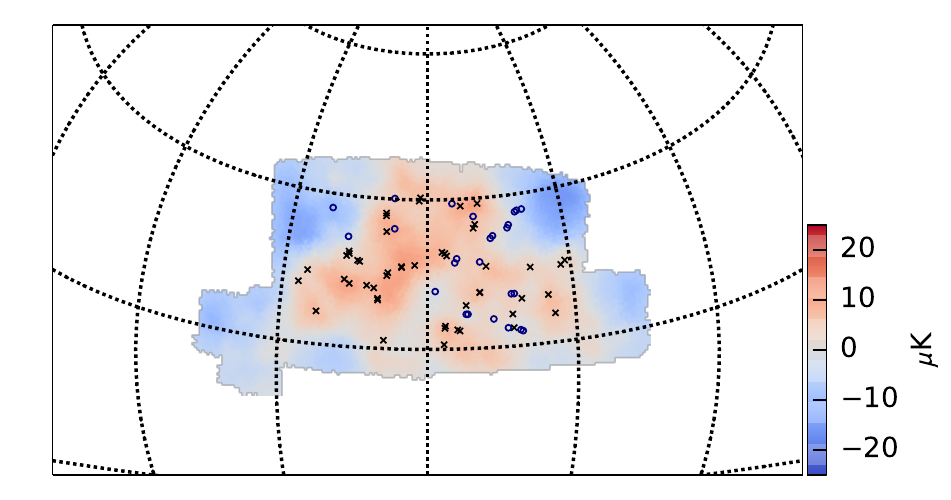}
\caption{Top row: CMB maps after low-pass filtering $\ell<20$.  Bottom row: ISW maps.  Cross symbols mark the angular posistion of local maxima of the gravitational potential and open circles mark local minima more than 10\degr{} from survey boundaries.  The locations of \citetalias{Granett08} supervoids and superclusters are indicated by the large open circles and crosses.  The CMASS survey boundary is overplotted in black over the CMB temperature maps. \label{fig:cmb} 
}
\end{center}
\end{figure*}

\begin{table}
\begin{center}
\begin{tabular}{llccc}
\hline
 Sample    &  & Number & Median z &  Galaxy bias     \\
\hline
$0.43<z<0.55$  & North   & 309900 & 0.503 &$1.94\pm0.01$ \\
                          & South   & 109793 & 0.499 & $1.97\pm0.01$ \\
\hline
$0.55<z<0.65$ & North    & 218311 & 0.590& $2.15\pm0.01$\\
                         & South    & 77872 &0.590& $2.18\pm0.02$\\
\hline
\end{tabular}\caption{Galaxy samples and best-fit bias parameters.  The bias parameters are derived after fixing the amplitude of the matter power spectrum with $\sigma_8=0.8$ under the fiducial cosmology given in the text. \label{tab:samples} 
}
\end{center}
\end{table}

\subsection{CMB maps}
The second data release from the Planck mission provides both temperature and polarisation maps \citep{Planck}.  We use the foreground cleaned SMICA temperature map downgraded to 1\degr{} resolution ($\nside=64$) \citep{PlanckMaps}.  The polarisation data may also be used to help separate the late-time and primordial temperature anisotropies.  However, as of the latest release, the polarisation data is not available on large scales, harmonic $\ell<40$, making the gain in the joint analysis negligible \citep{PlanckISW}.  Thus, we proceed with the analysis using only the temperature maps.

\section{ISW maps}\label{sec:iswmap}
A prediction of the linear ISW signal may be obtained from the galaxy density field by employing the Poisson equation \citep{Granett09}.
We use a Wiener filter to estimate the underlying matter density field and correct for survey completeness.   The Wiener filter gives the minimum-variance estimate
of the density field.  In the Bayesian formulation it corresponds to the maximum a-posteriori solution assuming a Gaussian likelihood and prior, assumptions that hold
for the galaxy field on large scales.

Our method is based on the Gibbs sampler developed to reconstruct the underlying density field of the VIPERS galaxy sample \citep{Granett15}
with adjustments to accommodate the CMASS sample.  We upweight the galaxies using the inverse of the completeness as described above.
 We divide the CMASS sample into two subsamples by redshift: $0.43<z<0.55$ and $0.55<z<0.65$.  The subsamples are combined to jointly estimate a single density field, but we allow the relative galaxy biases to vary.  The weighted number of galaxies from each subsample is  assigned over a comoving cartesian grid with cubic cells
$10\hmpc$ on a side.  We process separately the Northern and Southern Galactic caps.  The North is embedded into a cube with 256 cells on a side while the South fits in a slightly smaller box with size 200 cells.  The number of galaxies in a given cell indexed
by $i$ may be related to an underlying over-density field $\delta_g$,
\begin{equation}
N_i = w_i\bar{N}_i (1+\delta_{g,i}).
\end{equation}
The mean number density of the cell is $\bar{N}_i$ and the selection
function is $w_i$.  The selection function is estimated from the
normalised number of random points falling in the given cell.  It
includes the angular mask.  The redshift
distribution is contained in $\bar{N}_i$.  The expected shot noise in
the cell is $\sigma^2_i = \alpha w_i\bar{N}_i$, where $\alpha$ accounts for the completeness weights.

As in \citet{Granett15} our data model includes a constant galaxy bias  and the redshift distribution is left free.  However, in this analysis we fix the power spectrum to the fiducial model computed with CLASS to reduce the convergence time of the Gibbs sampler.  The nonlinear power spectrum is computed using Halofit \citep{Takahashi12} with the fiducial parameter set ($h=0.6774$, $\Omega_m=0.3089,\sigma_8=0.80$) at effective redshift $z_{ref}=0.55$).  Use of the non-linear power spectrum aids in accurately recovering small-scale structures.  Instead if we used the purely linear power spectrum the field would be unnecessarily smoothed.  We apply the Wiener filter isotropically in redshift space and so we use the monopole of the power spectrum estimated using the dispersion model with distortion parameters $\beta=0.375$ and $\sigma_v=400{\rm km/s}$
 \citep{PeacockDodds94,Cabre09}.

We find that the Gibbs sampler converges after 10 steps.  To produce realisations of the density field we ran 7 chains, each with 500 steps.  The constraints on the bias parameters are listed in Table \ref{tab:samples}.

From the density field we may estimate the gravitational potential using the Poisson equation in Fourier space,
\begin{equation}\label{eq:poisson}
k^2 \Phi(\vec{k}) = - \frac{3}{2}H_0^2 \Omega_m (1+z) \delta(\vec{k}).
\end{equation}
In doing so, we scale $\delta$ with the linear growth factor to a common epoch: $\delta(z_{ref}) = \delta(z)D(z_{ref})/D(z)$.
This computation is carried out via the FFT algorithm.

The ISW signal is an integral over the time-derivative of the potential along the line-of-sight,
\begin{equation}
T_{ISW} = -T_{CMB} \frac{2}{c^2} \int \frac{d\Phi}{d \tau} d\tau.
\end{equation}
However, this is simplified using the linear prediction, that is,
$d\Phi/d\tau \propto d (1+z)D(z)/d\tau$, to obtain,
\begin{equation}
\begin{split}
T_{ISW}(\theta,\phi) &=  -T_{CMB} \frac{2}{c^2}  \frac{1}{(1+z_{ref})D(z_{ref})} \\
& \times \sum_i  \Phi(\theta,\phi,r_i)  \left[ \frac{d (1+z)D(z)} {d r}\right]_{r_i} \Delta r
\end{split}
\end{equation}
which represents a projection over the potential field with a given kernel.
After projecting the potential along the line-of-sight of the North and South samples  we find the ISW maps shown in Fig. \ref{fig:cmb}.
In Appendix~\ref{app} we compare this map with the reconstruction from \citet{Granett09} which was based upon a photometric 
luminous red galaxy sample over a similar redshift range in the Northern field.

\subsection{Template fit}
The observed CMB temperature map may be written as a sum of components
including the primary temperature $T_{primary}$, the ISW signal
$T_{ISW}$ and the uncorrelated foreground contribution
$T_{foreground}$
\begin{equation}
T_{CMB} = T_{primary} + T_{foreground} +  \lambda T_{ISW}
\end{equation}

The maximum likelihood estimator for the amplitude parameter $\lambda$ is
\begin{equation}
\hat{\lambda} = \frac{T_{CMB} C^{-1} T_{ISW}}{T_{ISW} C^{-1} T_{ISW}}. \label{eq:tempfit}
\end{equation}
The variance is given by
\begin{equation}
\sigma^2 = \left(T_{ISW} C^{-1} T_{ISW}\right)^{-1}.\label{eq:temperr}
\end{equation}

The covariance matrix $C$ is given by $C=\langle
(T_{primary}+T_{foreground})^2\rangle$.  We use the best fitting
temperature power spectrum model from Planck computed using CLASS.  In
pixel space, the covariance matrix is
\begin{equation}
C_{ij} = \sum_l\frac{2l+1}{4\pi} C_l W_l^2P_l(\cos\theta_{ij}),
\end{equation}
where $P_l$ is the Legendre polynomial and $\theta_{ij}$ is the
angular separation between two healpix cells indexed by $i$ and $j$.
The window function $W_l$ is set by the Healpix pixelization at
$\nside=64$.  At this resolution we neglect the beam and instrument
noise. 

A correction to the covariance must be added to address the fact that
we estimate the mean of the distribution from the sample itself.  The
covariance is $C_{ij}'=\langle (x_i-\xbar) (x_j-\xbar) \rangle$ where
$\xbar$ is the mean pixel value within the survey:
$\xbar=\frac{1}{N_{pix}}\sum_i x_i$.  Expanding, this becomes
$C_{ij}' = \langle x_i x_j \rangle - \frac{1}{N}\sum_\alpha \langle
x_i x_\alpha\rangle - \frac{1}{N}\sum_\alpha \langle x_\alpha x_j
\rangle + \frac{1}{N^2}\sum_\alpha \sum_\beta \langle x_\alpha x_\beta
\rangle.$ 

The results of the template fit are presented in Table \ref{tab:fit}.  Considering the Northern map, we find that the ISW amplitude
 $\lambda=0.41$ with signal-to-noise 2.1.  The Southern map gives no indication of a signal with an amplitude consistent with zero.
In combination, we find $\lambda=0.33\pm0.17$ giving a one-tailed p-value $p=0.03$.  

\subsection{Discussion}
The template fit amplitude  gives the degree of agreement between the fluctuations observed on the CMB and those expected from the galaxy field through the linear ISW effect.  If the ingredients of the model are correct we expect an amplitude of unity.  The value less than 1 means that we have over-predicted the fluctuations on the CMB.
Physically we may relate this to the  amplitude of galaxy fluctuations through the bias parameter implying a galaxy bias larger than anticipated.  The amplitude $\lambda=0.4$ may be reconciled by increasing the bias by a factor of 2.5.  

Previously, \citet{Granett09} found a template fit amplitude greater than 1.  In that work, it was found that the amplitude depended
on the number density, redshift distribution and redshift uncertainty of the galaxy tracers and furthermore, that shot noise biases the amplitude low.  Thus at this point, we do not draw 
conclusions from the value of the amplitude and take the signal-to-noise ratio as confirmation of the validity of the map.

\begin{table}
\begin{center}
\begin{tabular}{lrrrr}
\hline
\multicolumn{5}{c}{$\Delta T_{ISW}$ Template fit}\\
Field   & $\lambda$ & $\sigma$ & $\lambda/\sigma$  & 1-tail $p$\\
\hline
North & 0.41 & 0.20 & 2.1 & 0.02 \\
South & 0.09 & 0.34 & 0.3 & 0.4 \\
Combined & 0.33 & 0.17 & 1.9 & 0.03 \\
 \hline
\multicolumn{5}{c}{$C^{gT}_\ell$ Power spectrum fit} \\
Field   & $A$ & $\sigma$ & $A/\sigma$  & 1-tail $p$\\
\hline
North & 2.5 & 1.1 & 2.2 & 0.01 \\
South & 2.1 & 1.9 & 1.1 & 0.1 \\
Combined & 2.4 & 1.0 & 2.4 & 0.01 \\
\hline
Clean North & 0.7 & 1.1 & 0.6 & 0.3 \\
Clean South & 1.8 & 1.9 & 0.9 &  0.2\\
Clean Comb. & 0.9 & 1.0 & 0.9 & 0.2 \\
 \hline
 
\end{tabular}\caption{The best-fit model template amplitudes for the ISW map and angular power spectrum measurements. \label{tab:fit}}
\end{center}
\end{table}

\begin{figure*}
\begin{center}
\includegraphics[width=7in]{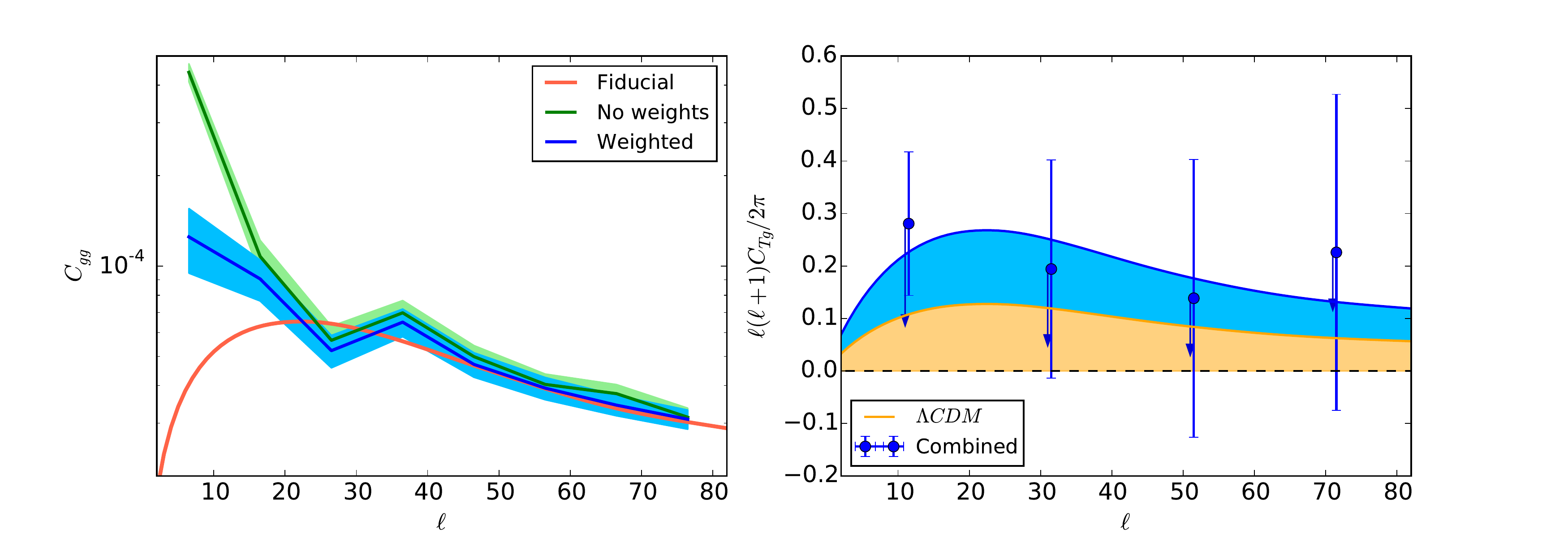}
\caption{Left: the angular auto-power spectra of BOSS galaxies.  The markers with error bars give the mean value in a bin of width $\Delta \ell=10$. We show the measurement with and without the corrective weights for foreground systematics.  Over-plotted with the red solid curve is a fiducial power spectrum computed with CLASS and halofit.  We find that the data is consistent with a bias $b_g=2.0$.  Right, the cross galaxy-CMB temperature power spectrum.  The fiducial model is over-plotted and we find a best fitting amplitude $\lambda=2.4\pm 1.0$.  After
subtracting the ISW map from the CMB, the cross-correlation drops, shown by the arrows.}
\label{fig:cl} 
\end{center}
\end{figure*}
\section{Angular cross-correlation}\label{sec:crosscorr}

A standard probe of the ISW effect is the cross-correlation of galaxy density fields and CMB temperature maps. For SDSS luminous red galaxy data sets, previous studies found a moderately significant $2-2.5\sigma$ correlation with amplitudes higher than expected in the $\Lambda CDM$ model \citep{Ho08,Giannantonio08,Granett09}. Other studies reported ISW signals consistent with $\Lambda CDM$ \citep{HM14,Giannantonio14}.

We select CMASS galaxies in the redshift range $0.43<z<0.65$ and compute the projected density field on a Healpix grid with resolution 1\degr{} ($\nside=64$).
The selection function is estimated by assigning particles from the random catalogue in the same fashion.  We then normalise the galaxy count map by the density randoms in each cell, setting to zero cells that are less than 50\% complete.  

\subsection{Auto power-spectra}\label{sec:autocorr}

We compute harmonic-space power spectra, $C_l^{XY}=\langle a_{lm}^X a_{lm}^Y\rangle$, using the SpICE code \citep{spice} which properly accounts for the survey geometry.
We first compute the auto-power spectrum of this map as a function of spherical harmonic index $\ell$, $C^{gg}_\ell$, shown in Fig. \ref{fig:cl}.
We measure consistent values in our bins of width $\Delta \ell=10$ in the Northern and Southern hemispheres.  
We thus analyse multipoles $\ell < 80$ since the theoretical S/N for an ISW signal in the $\Lambda CDM$ model already converges at $\sim1.7$ for $\ell \approx 60$ \citep{HM14}.
For comparison, we overplot a model computed under best-fit Planck cosmological parameters ($h=0.6774$, $\Omega_m=0.3089,\sigma_8=0.80$) at effective redshift $z_{ref}=0.55$. The constant galaxy bias $b_g=2$ provides a good fit.
The systematic weights alleviate the excess power on large-scales.

The fiducial model is given by the projection of the power spectrum in the plane-parallel limit:
\begin{equation}
C_l^{gg}= \int dr\ r^2 n(r)^2 \ b_g^2(r)\  \frac{D^2(r)}{D^2(z_{ref})}\ P\left(\frac{l+1/2}{r} | z_{ref}\right).
\end{equation}
The redshift distribution is normalised such that
$\int dr\ r^2 n(r) = 1$ , $b(r)$ is the galaxy bias, $D(r)/D(z_{ref})$ is the growth factor that linearly evolves the field to redshift $z_{ref}$ and $P$ is the non-linear power spectrum estimated at redshift $z_{ref}$.  We assume that the galaxy bias is constant over the redshift range of interest.

\subsection{Cross power-spectra}

Similarly, the expected angular power spectrum in the flat-sky limit, see 
 \citet{Afshordi04}, is an integral over the line-of-sight distance
$r$,
\begin{eqnarray}
C_l^{\delta T} &=& \tcmb\frac{3 H_0^2 \Omega_m b_g}{c^2} \frac{1}{(l+1/2)^2} \nonumber \\
&&\times \int dr\ r^2 n(r) \frac{d(1+z)D_1(z)}{dr} P\left(\frac{l+1/2}{r}\right),
\end{eqnarray}
where, $n(r)=\frac{dN(r)}{dz dV}$ is the galaxy selection function
normalized by $\int r^2 \frac{dN}{dz dV}dr$, $D_1$ is the growth
factor and $P\left(k=\frac{l+1/2}{r}\right)$ is the matter power spectrum at
redshift $z_{ref}$.

In Fig. \ref{fig:cl} we show the cross correlation between the galaxy density field and the CMB.  We allow the amplitude of the model to be free with parameter $A$ and find the best-fit value and uncertainty.  We introduce  the amplitude parameter $A$ which has a distinct interpretation from the amplitude $\lambda$ used in the template map fit.  The value $A=1$ would imply consistency with the \LCDM{} expectation with the best-fit galaxy bias and $A>1$ can be reconciled by increasing the galaxy bias. The fit results are summarised in Table \ref{tab:fit}. While systematic differences between the two hemispheres may affect our measurement \citep{Giannantonio14}, we find consistent ISW signals for the Northern, Southern, and combined maps.

As a consistency test of the ISW map derived in Sec. 3, we subtract it from the Planck CMB temperature map and refer to this as the ISW cleaned map.  
Repeating the cross-correlation measurement, we find that the amplitude drops to $A=0.9\pm 1.0$ when the galaxy density map is cross correlated with the ISW cleaned CMB map. The match between the two methods is better for the Northern field, while a residual remains in the South at the 1$\sigma$ level.  

\begin{figure*}
\begin{center}
\includegraphics[scale=0.9]{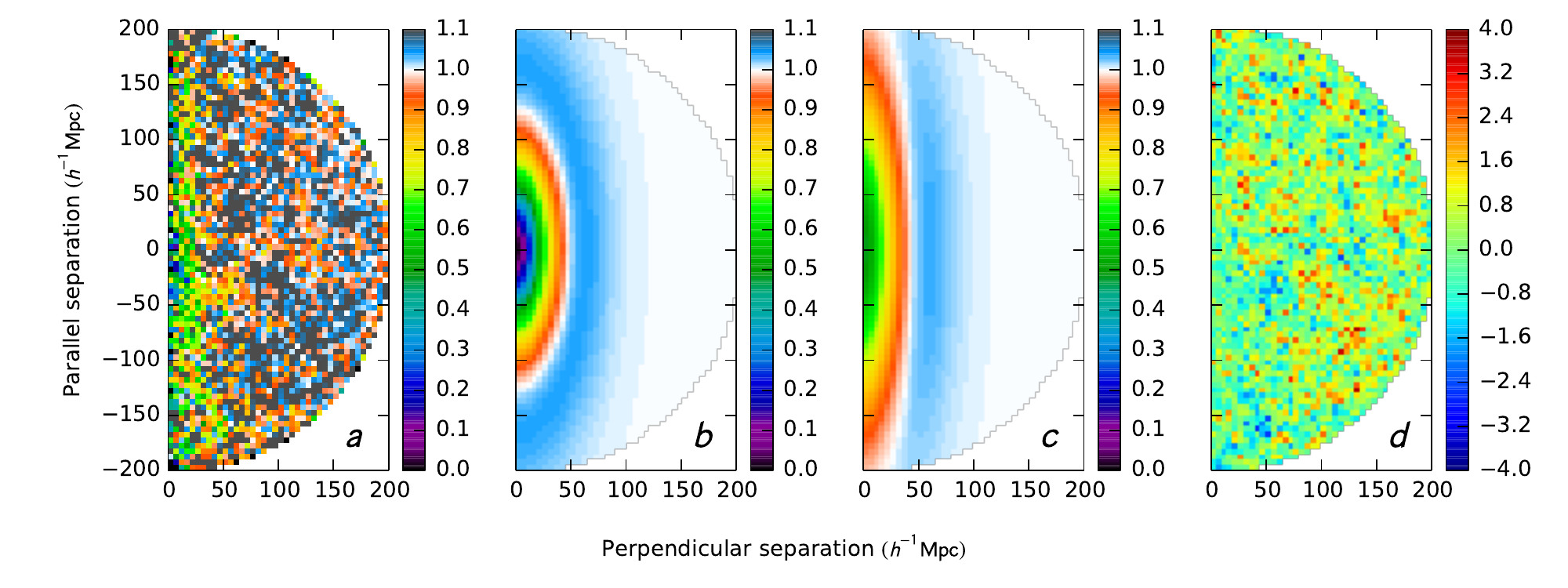}
\caption{Mean \citetalias{Granett08} supervoid profile:  (a) the normalised number density of CMASS galaxies measured in bins perpendicular and parallel to the line-of-sight (bin size is $5\mpch$), (b) the best-fit model density profile of the void, (c) the model convolved with the best-fit Gaussian line-of-sight position error showing how it would appear in observations, (d) the residual in units of standard deviation after subtracting the convolved model from the data.  \label{fig:void2d}}
\includegraphics[scale=0.9]{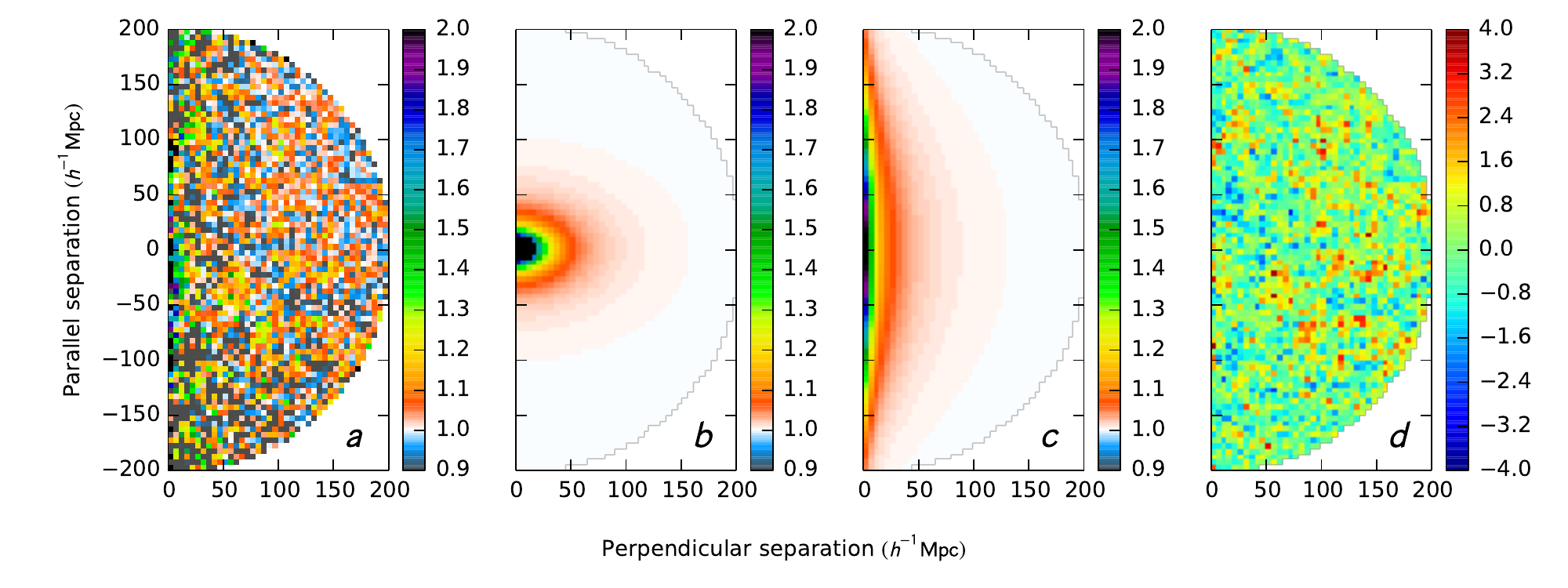}
\caption{Same as Fig. \ref{fig:void2d}, but for the \citetalias{Granett08}  supercluster catalogue. \label{fig:clus2d}}
\end{center}
\end{figure*}

\begin{table}
\begin{center}
\begin{tabular}{cccc}
\hline
\multicolumn{4}{c}{Supervoids}\\
 $r_v\  (\mpch)$ &   $\alpha$ &  $q$ & $\sigma_{los}\  (\mpch)$  \\
\hline
$35.9 \pm 3$ & $1.7\pm0.15$   &   $2.6 \pm 0.4$ & $76 \pm 9$ \\ 
\hline
\multicolumn{4}{c}{Superclusters}\\
 $r_s\  (\mpch)$ &   $\delta_c$ &  $q$ & $\sigma_{los}\  (\mpch)$  \\
\hline
$9.0 \pm 3$ & $167^{+220}_{-85}$   &   $0.64^{+0.3}_{-0.2}$ & $79 \pm 9$ \\ 
\hline
\end{tabular}\caption{Best fit parameters for \citetalias{Granett08} supervoid and supercluster mean profiles with 68\% confidence limits.
 }\label{tab:fit2}
\end{center}
\end{table}

\section{Characterisation of \emph{GraNeSz} super-structures}\label{sec:radprof}
\subsection{Supervoid and supercluster profiles}
The spectroscopic CMASS sample allows us to measure the shape and sizes of \citetalias{Granett08} super-structures which were originally identified in a photometric redshift sample.
We begin by measuring the  number density of CMASS galaxies around the positions of \citetalias{Granett08} super-structures.   The catalogue includes 50 supervoids and 50 superclusters.
We measure the mean density profile by taking the  ratio of the number of galaxies to the normalised number of unclustered random points in a given separation bin.  The number of points in the  random catalogue is  17 times the total number of gsalaxies.
Fig. \ref{fig:void2d} and \ref{fig:clus2d} show the mean number density of the 50 supervoids and 50 superclusters as a function of distance parallel and perpendicular to the line-of-sight.   We find that the profiles are not localised in the line-of-sight direction.  This may be due to error in the individual centres estimated with photometric redshifts.  To interpret these plots we will fit simple void and cluster profile models that includes uncertainty in the structure centre.

We estimate the variance of the profile measurements by generating many realisations centred on random positions in the survey.  First, considering either the supervoid or supercluster catalogue, 50 random points are drawn from a random catalogue over the Northern field survey volume.  Redshifts are assigned from the true redshifts in the superstructure catalogue.  Then, the mean profile is measured from these positions.    We used 1000 realisations of the process to estimate the diagonal elements of the covariance matrix for both supervoid and supercluster catalogues.

\begin{figure}
\begin{center}
\includegraphics[scale=0.7]{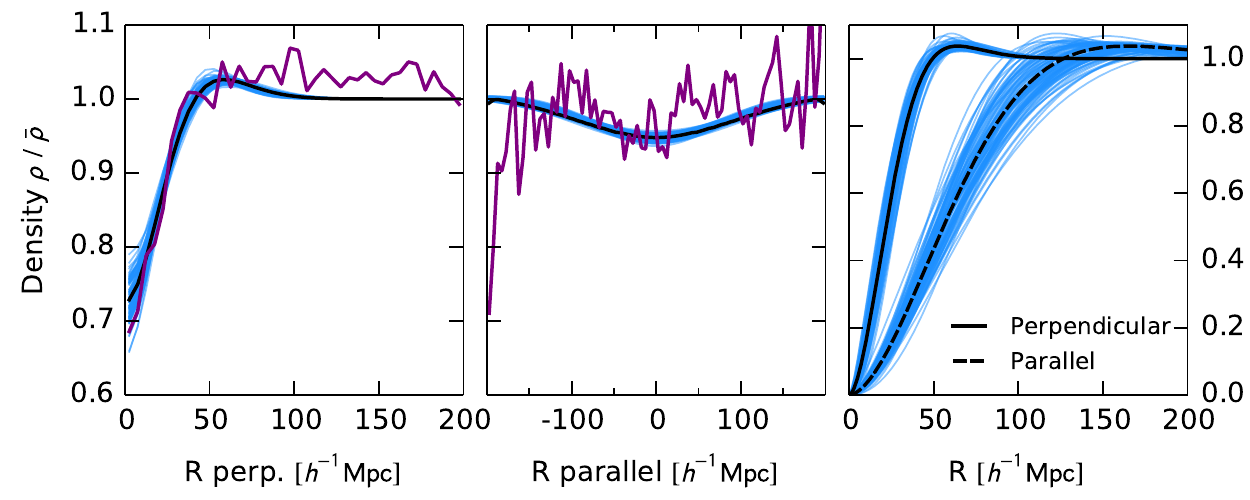}
\caption{The \citetalias{Granett08} supervoid profile (left) projected parallel to the line-of-sight, (center) projected perpendicularly to the line-of-sight, (right) the deconvolved three-dimensional model.  The measurements are compared with the best-fit model (dark solid curve) and samples taken from the Markov chains (light blue curves).
\label{fig:voidmodel}}
\includegraphics[scale=0.7]{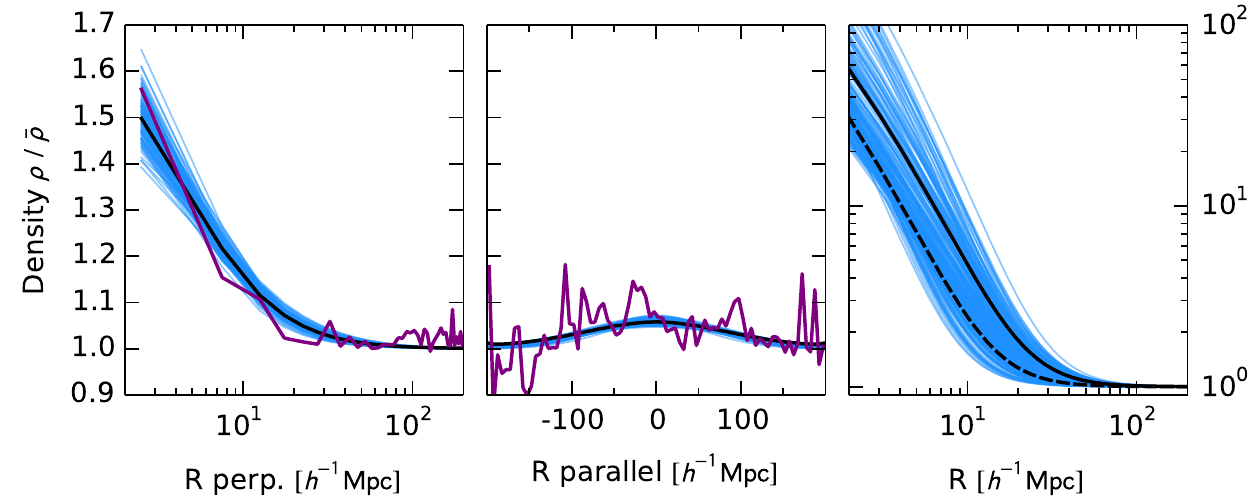}
\caption{The same as Fig. \ref{fig:voidmodel}, but for \citetalias{Granett08} superclusters.  \label{fig:clusmodel}}
\end{center}
\end{figure}

\begin{figure}
\begin{center}
\includegraphics[scale=0.9]{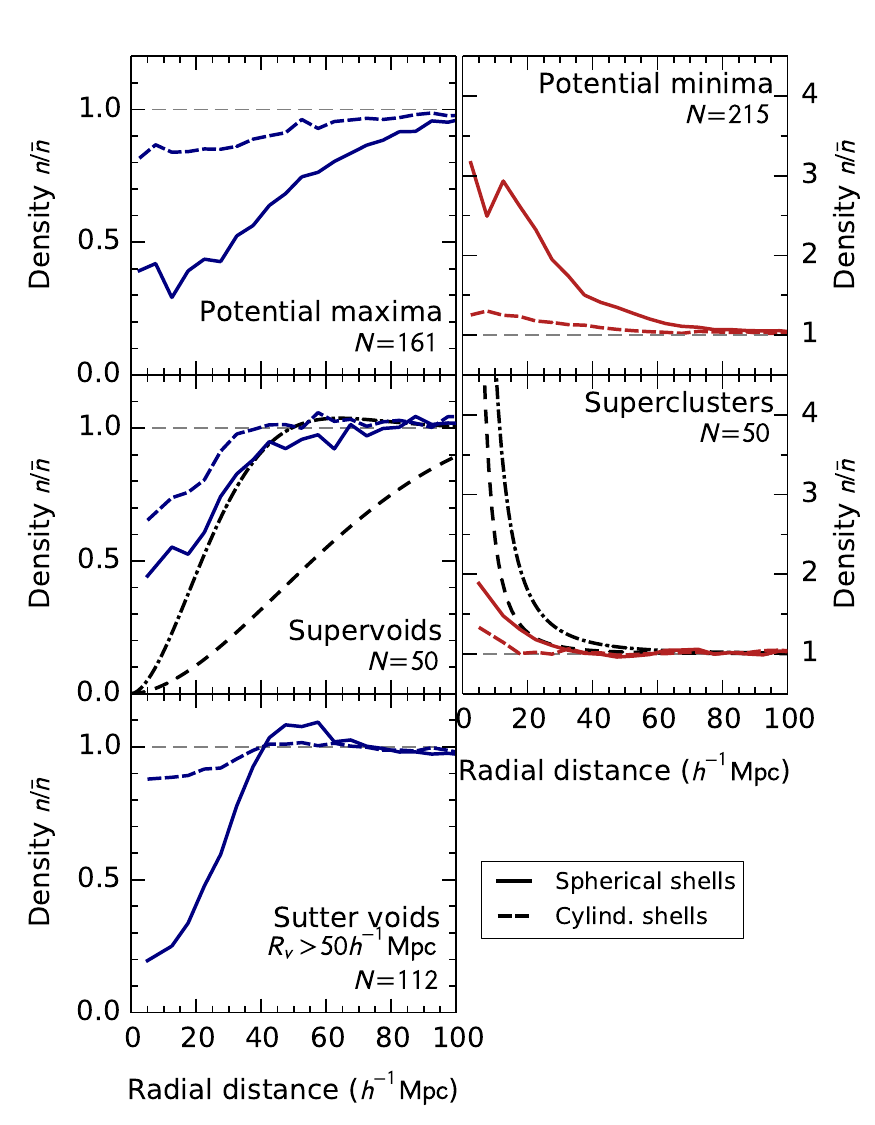}
\caption{Radial density profiles of BOSS super-structures.  Top row: local minima and maxima of the gravitational potential. Middle row: \citetalias{Granett08} supervoids and superclusters.  The best-fit models for the three-dimensional profile parallel (long-dashed curve) and perpendicular (dot-dashed curve) to the line-of-sight are shown.   Bottom row: supervoids from the Sutter DR10 analysis. Plotted are profiles in spherical shells as well as cylindrical shells along the line of sight within  $r=200\mpch$  to show a representation of the structure in photometric redshift space.\label{fig:prof}}
\end{center}
\end{figure}

\subsection{Void and cluster models}\label{sec:fit}

We use the following analytic form for the void profile with scale radius $r_v$ and shape parameter $\alpha$ (Hawken et al, {\it in prep.}):
\begin{equation}
\frac{\rho(\tilde{ r})}{\bar{\rho}} = 1 - \left(1 - \frac{\alpha \tilde{ r} ^\alpha}{3r_v^\alpha}  \right) \exp\left[-\left( \frac{\tilde{ r}}{r_v}\right)^\alpha\right].
\end{equation}
The void is permitted to be elongated along the line-of-sight with ellipticity parameter $q$, and we define:
\begin{equation}
\tilde{ r}^2 = r_{perp}^2 + r_{par}^2/q^2.
\end{equation}

Superclusters are known to have irregular morphologies \citep{Einasto14} and may not be well approximated by spherical symmetry.  Nevertheless we apply the 
  \citet[][NFW]{Navarro96} analytic form to characterise the mean supercluster profile:
\begin{equation}
\frac{\rho(\tilde{ r})}{\bar{\rho}} = 1 + \frac{\delta_c}{\tilde{r}\left(1+\tilde{r}/r_s\right)^2}
\end{equation}
The supercluster profile is also allowed to be elongated along the line-of-sight with parameter $q$.

We allow for a scatter in the line-of-sight position by introducing a convolution with a Gaussian kernel characterised with scale $\sigma_{los}$:
\begin{equation}\label{eq:conv}
\rho_z(r_{perp},r_{par}) = \frac{1}{\sqrt{2\pi \sigma_{los}^2 }} \int  e^{-\frac{(y-r_{par})^2}{2\sigma_{los}^2}} \rho(r_{perp},y) dy.
\end{equation}

We fit the model parameters to the measurements using a Markov chain monte carlo method implemented with the Emcee package\footnote{\url{http://dan.iel.fm/emcee/}} \citep{emcee}.  The fit is carried out using the diagonal covariance matrix constructed as described above.

We give the medians of the parameter chains along with the 68\% confidence levels in Table \ref{tab:fit2}.   In Figs. \ref{fig:void2d} and \ref{fig:clus2d} we show the mean anisotropic profiles of the voids and clusters along with the best-fit models with their residuals.  The fit is carried out on the model convolved with the Gaussian kernel (Eq. \ref{eq:conv}) but we also show the profile before the convolution step to represents the underlying galaxy distribution.
Figs. \ref{fig:voidmodel} and \ref{fig:clusmodel} compare the model with perpendicular and parallel projections of the profiles along with the model three-dimensional radial profile.  The goodness of fit is $\chi^2_{void}=0.93$ and $\chi^2_{cluster}=0.99$ per degree of freedom for the supervoid and supercluster models respectively.

In principle we could recenter each structure along the line-of-sight to correctly superpose the profiles.  However, we found that due to the sparseness of the spectroscopic catalogue, the true centres could not be accurately estimated and so the uncertainty in the position can only be corrected for statistically on the mean profile.

\subsection{Discussion}

The model fits demonstrate that  the supervoid and supercluster samples both represent large structures.  The voids reach to $R=50\mpch$ and the scale radius for superclusters is $r_s=9 \mpch$.

The mean supercluster profile is much broader than that which would be expected for a typical massive cluster.  We find that the radius that encloses a mean density $200\rho_{crit}$ is $r_{200}=1.1 \mpch$ giving a concentration index $c_{200}\equiv r_s/r_{200}$ of $c_{200}=  0.12$ while massive clusters have concentration $c_{200}\approx 3$ \citep{Umetsu15}.  Given the low concentration index, the mass within $r_{200}$ is modest: we find $M_{200}=1.5\times 10^{14} h^{-1} {\rm M_{\odot}}$.  If we instead integrate to the scale radius $r_s$ we find $M(r<r_s)=4.7\times10^{15} h^{-1} {\rm M_{\odot}}$.  For a typical massive cluster these masses are $M_{200} =1\times10^{15} h^{-1} {\rm M_{\odot}}$  and $M(r<r_s)=2\times10^{14} h^{-1} {\rm M_{\odot}}$.  The scale of the superclusters suggests that they are not virialized structures.  Despite this, we find that the NFW profile characterises the mean profile well.

The best-fit void model is elongated along the line-of-sight with axis ratio $q=2.6 \pm 0.4$.  We found that a spherically symmetric model is unable to fit the data even with the free line-of-sight smoothing parameter.  We explored this peculiarity with more general void profile forms, but found that even a spherically symmetric step function profile is unable to reproduce the measurement.    On the other hand, the supercluster model is not strongly elongated: we find only a marginal squashing with $q$ discrepant from unity at the $1\sigma$ level.

How can we interpret the apparent elongation of the supervoids?  This is likely to be a real effect arising from the void-finding algorithm.  \citetalias{Granett08} identified the supervoids and superclusters in a photometric redshift catalogue with redshift error $\sigma_z\sim0.05\sim 100\mpch$.  We suggest that when applied to a photometric sample the algorithm is most sensitive to under-dense tubes that are pointed along the line-of-sight.  The effect is important  for voids because they are large, irregularly shaped structures.  Superclusters are also known to  have wall-like or filamentary morphologies \citep{Einasto14}; however, we find that the structures in the supercluster catalogue do not have a preferred alignment. 

The void and cluster models incorporate the position uncertainty through the smoothing parameter $\sigma_{los}$.  We found that the parameter is  required to fit both the supervoid and supercluster measurements; moreover the scales determined from the two models are in excellent agreement.  Interpreting this effect as arising from redshift error, the scale  $\sigma_{los}=76 \mpch$  corresponds to a redshift error of $\sigma_z=0.034$.

\section{Comparison of BOSS super-structures}\label{sec:super}

\subsection{Local extrema of the gravitational potential}
We  construct a new superstructure catalogue defined by the positions of local extrema of the gravitational potential field.  
The starting point is the Wiener estimate of the gravitational potential derived through the Poisson equation, Eq. \ref{eq:poisson}.
We perform the computation on a grid with cell size $10\mpch$, thus we may resolve structures on scales $R>20\mpch$.  To find the
largest structures, we convolve the field with a three-dimensional Gaussian filter with full-width half-maximum $60\mpch$.  
The local extrema of the field then correspond to cells that are more extreme, either greater or less than, their six adjacent cells.  
 The positions of the extrema are discretised by the grid.  This introduces an uncertainty  of  $10\mpch$ which corresponds to an angular scale of $\sim 30$ arcmin; however, this uncertainty in the structure positions is much less than the smoothing scale.
Minima of the potential correspond to over-densities, while maxima correspond to under-densities in the galaxy field. We exclude objects with centres less than 10\degr{} from the survey boundary.
 We find 136 and 25 supervoids in the North and South, respectively, and 166 and 49 superclusters in the North and South, marked in Fig. \ref{fig:cmb}.

\subsection{Radial density profiles}
In Fig. \ref{fig:prof} we compare the radial density profiles of three super-structure catalogues: \citetalias{Granett08}, peaks and troughs of the potential and the DR10 spectroscopic void catalogue by \citet{Sutter14}.   The projected profiles are computed by averaging in cylinders along the line-of-sight within a sphere of $r=200\mpch$.  
 In the case of the \citetalias{Granett08} structures the void centres are given by the photometric redshift estimates, but we overplot the deconvolved best-fit model found in Sec. \ref{sec:fit}.   The Sutter voids were selected in the redshift range $0.5<z<0.6$ and have radii $R>50\mpch$.   We found that for voids with radius less than $40\mpch$ the structures in projection are overdensities, a fact pointed out by \citep{Nadathur14,Cai14}.   We find that the GraNeSz voids are deeper in projection than the Sutter voids, reinforcing the fact that  photometric and spectroscopic void finders can be sensitive to different types of structures. 

The superstructures identified from local extrema of the gravitational potential appear significantly larger in extent.  The supervoids appear significantly underdense, deeper than \citetalias{Granett08} supervoids and match the depth of \citet{Sutter14} catalogue, but lacking the compensated ring structure.  Moreover, in projection they appear deeper than the \citetalias{Granett08} voids. The superclusters are found to be significantly more dense than \citetalias{Granett08} superclusters, both in three-dimensions and projection.

\subsection{Radial temperature profiles}
\begin{figure}
\begin{center}
\includegraphics[scale=0.9]{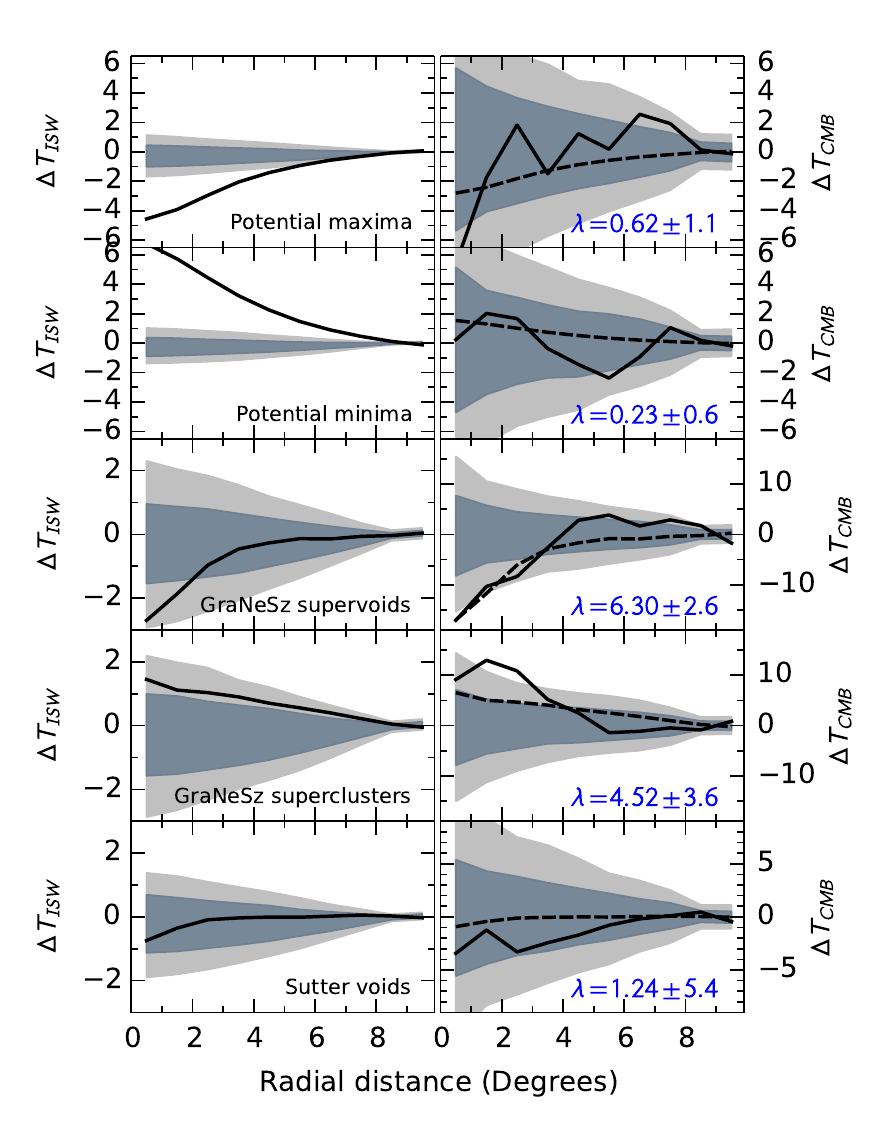}
\caption{ Mean temperature profiles of super-structure catalogues.  Left: the profile of structures projected on the ISW map; right:  the measured temperature profile on the CMB.  The shaded regions show 68\% and 95\% confidence intervals.
   \label{fig:iswprof}}
\end{center}
\end{figure}

We now examine the imprint of these structures in our model ISW map and in the Planck temperature map.  We will use the ISW map as a template to  match to the CMB data.
The radial ISW temperature profiles are shown in the left column of Fig. \ref{fig:iswprof} and the temperature profiles on the CMB are shown on the right.  The shaded regions show the 68 and 95\% confidence intervals.  For the ISW map, the confidence interval was determined by measuring the profile around randomly distributed points within the survey mask.  We use 1000 random realisations to compute the variance of the profile in this way. On the right, for the variance on the CMB temperature, we keep the coordinates of the structures fixed and generate mock CMB skies using the Healpix Synfast code.  Here we also use 1000 realisations and compute the covariance matrices for the profiles.

The amplitude of the profile is affected by long-wavelength fluctuations that effectively modulate the background level.  To remove this variance we apply a high-pass filter to both the ISW and CMB maps.  We use a filter with a cosine cut-off at $\ell=10$, corresponding to 20\degr{} scales.  In presenting the profile, we further subtract the mean temperature at 10\degr{} scales.  We do not expect such filtering to affect a well-localised signal arising from individual voids or clusters.

We see that each super-structure sample is correlated with the ISW map to some degree.  The peaks and troughs of the potential have the largest amplitude, which can be expected by their construction.  The \citetalias{Granett08} sources do not correspond to the most extreme features in the map, but show significant correlation.  On the other hand, the \citet{Sutter14} sample shows only a weak correlation with the ISW map, which may be anticipated by their shallow profiles in projection.

We use the ISW profile measured on the ISW map as a template to fit the CMB temperature profile.  The template fit statistic and variance are computed as given by Eqs. \ref{eq:tempfit} and \ref{eq:temperr}.  The resulting significances are listed in Table \ref{tab:super}.  We find marginal detections of the \citetalias{Granett08} supervoids and superstructures.  The supervoid sample gives higher significance than the supercluster because there is better agreement between the measured profile and the template shape.
For the other samples the fits are consistent with no correlation.

\begin{table}
\begin{center}
\begin{tabular}{lrrrrr}
\hline
Sample   & $N$ & $\lambda$ & $\sigma$  & $\lambda/\sigma$ \\
\hline
GraNeSz supervoids     & 50 &  6.30   &  2.59    & 2.4      \\
GraNeSz superclusters & 50 &   4.51  &  3.57    & 1.3     \\
 \hline
Potential minima & 215 & 0.23 & 0.59 & 0.4 \\
Potential maxima & 161 & 0.61 & 1.05 & 0.6 \\
\hline
 Sutter DR10      & 112 & 1.23  & 5.39 & 0.2 \\
 \hline
\end{tabular}\caption{ISW source template fit significances.   \label{tab:super}}
\end{center}
\end{table}

\section{Conclusions}\label{sec:concl}
The BOSS survey volume and sampling rate allow a detailed investigation of ISW sources at redshift $z=0.5$.  We have used the spectroscopic sample to further investigate the \citetalias{Granett08} super-structure catalogue and carried out a preliminary analysis of public catalogues.  Our main findings may be summarised as follows.

\begin{itemize}
\item We constructed a map of the ISW anisotropy sourced by CMASS galaxies and confirmed that the signal is present in the CMB temperature map at the 97\% confidence level.  The detection is not significant in the Southern field.  This may be due to the small area and intrinsic variance of the signal or contamination by strong foreground systematics in the South.
\item We demonstrated that the signal is consistent with the measured galaxy density - temperature cross-correlation function with similar North-South differences.
\item We carry out a detailed characterisation of the \citetalias{Granett08} super-structures.  We confirm that they are indeed vacuous supervoids and extensive supercluseters, but they are not well localised in the line-of-sight direction.  We model this uncertainty as a convolution of the  profile.  Fitting the model to the observed mean profile, we find that the supervoids are intrinsically elongated along the line-of-sight.  We suggest that this feature arises from the void-finding algorithm run on a photometric redshift sample. 
\item We construct a new super-structure catalogue based upon peaks and troughs of the gravitational potential.  These structures have a much larger extent that the  \citetalias{Granett08} and \citet{Sutter14} objects, and should be expected to give a strong linear ISW contribution.
\item We compare the mean radial density profiles of structures and find that \citet{Sutter14} voids are relatively shallow in projection owing to the compensated shell.  The voids found with photometric redshifts do not show this morphology.
\item We construct ISW signal templates for the superstructure catalogues based upon our ISW map.  We use the template as a filter on the CMB to measure the ISW contribution, largely free of  `look elsewhere' bias.  With this method we find marginal detections of the \citetalias{Granett08}  super-structures, but the other super-structure samples are consistent with no signal.
\end{itemize}

We have resolved remaining questions regarding the nature of the  \citetalias{Granett08} super-structures by quantifying their morphologies; however, their apparent imprint on the CMB remains unexplained.  The elongation of the voids may offer a clue: we expect that the extreme tail of the ISW signal is sourced by structures that happen to be elongated along the line-of-sight \citep{Flender13}.  However, the linear ISW signal that we estimate based upon the reconstructed potential  does not account for the  measured temperature anisotropy.   Moreover our analyses of other super-structure catalogues give results fully consistent with the expectation in the \LCDM{} model.  Thus we are unable to confirm the existence of anomalous ISW sources.

 The approach we have developed here is fully empirical: we can characterise the sources expected to contribute most strongly to the ISW signal.  However there is also an important need for theoretical expectations  that can be provided by N-body simulations.  The picture may come into focus with future refinements of super-structure catalogues in BOSS.  We also look to upcoming  photometric and spectroscopic surveys to improve the fidelity of the galaxy density field reconstructions on large scales.

\appendix 
\section{Comparison with photometric ISW map} \label{app}
\begin{figure}
\begin{center}
\includegraphics[scale=0.9]{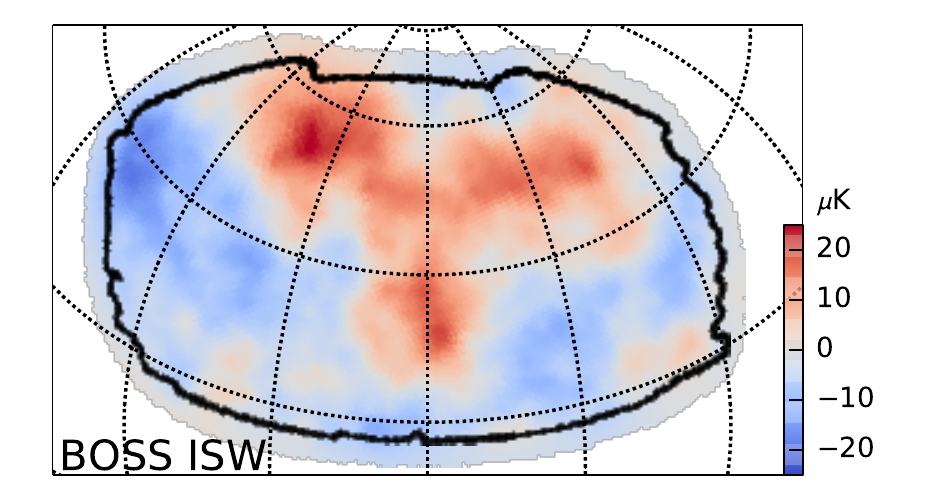}
\includegraphics[scale=0.9]{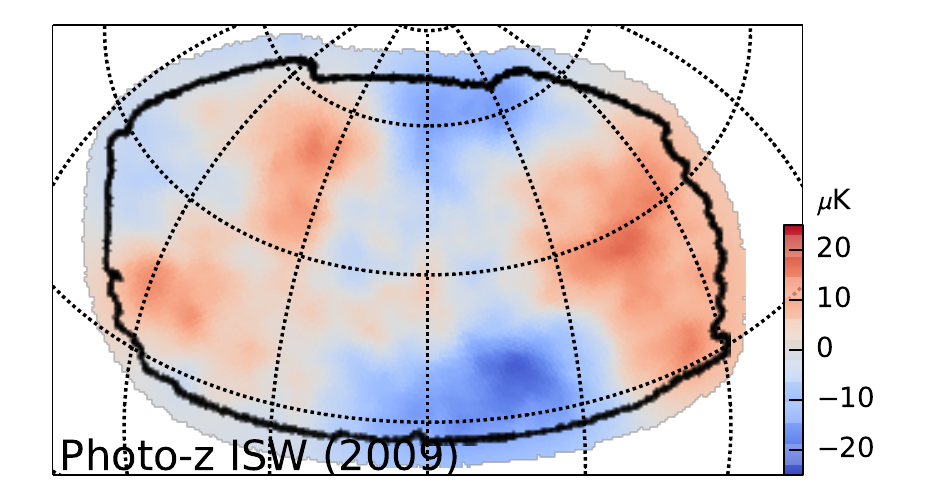}
\includegraphics[scale=1.0]{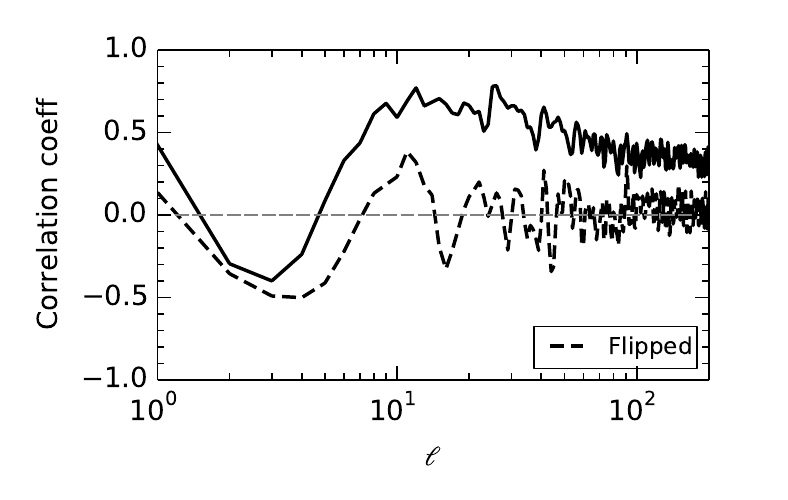}
\caption{ A comparison of the ISW map derived in this work (top frame) with the map estimated using a photometric sample of luminous red galaxies from \citet{Granett09} (middle frame).  The correlation coefficient between the two maps  is plotted in the bottom frame as a function of harmonic multipole $\ell$ (solid line).  The dashed line shows the result after flipping one of the maps in right ascension.
   \label{fig:compisw}}
\end{center}
\end{figure}

We have probed the ISW signal over the same area of sky and in a similar redshift range as an earlier work by \citet{Granett09}.  This previous study used a photometric sample of luminous red galaxies in the redshift range $0.4<z_{\rm phot}<0.6$ based on photometric redshifts.  Since the integrated linear signal depends on the redshift of the sources only weakly through the growth factor we expect that
maps generated with photometric or spectroscopic redshifts should be equivalent.
The  maps from the two works are shown in Fig. \ref{fig:compisw}.  By eye we can find some correspondence between the two maps for instance the hot spot in the top left  at RA, dec. 210\degr{}, 50\degr{} is in common; however, the large-scale fluctuations do not appear to match up.  

For quantitative comparison we compute the correlation coefficient of the spherical harmonic decompositions of the two maps $A$ and $B$ indexed by $\ell$: $r_{\ell} = C_{\ell}^{AB}/\sqrt{C_{\ell}^A C_{\ell}^B}$ plotted in the bottom panel of Fig. \ref{fig:compisw}.  On large scales we find an anti-correlation with $r<0$.  The correlation becomes positive at $\ell>10$ (20\degr{} scales) with $r\sim0.6$.  The mean correlation between the two maps is $r=0.2$.  As a sanity-check we flip one of the maps in right ascension and repeat the measurement.  This results in a reduced correlation coefficient as shown by the dashed curve in the figure.

We find that the two maps are correlated on small scales at the 60\% level but it is not perfect.  This may be explained by the noise properties of the samples: the photometric survey has a greater density of galaxies and so gives an estimate with reduced shot noise.  Furthermore, the redshift distributions of the samples do differ.  


\section*{Acknowledgments}
{\small
We thank Simone Aiole, Seshadri Nadathur, Mark Neyrinck and Istv\'{a}n Szapudi for discussing this work with us.
BRG and AJH acknowledge
support of the European Research Council through the Darklight ERC
Advanced Research Grant (\# 291521).
AK was supported by the Spanish Ministerio de Econom\'ia y Competitividad (MINECO) under projects FPA2012-39684, and Centro de Excelencia Severo Ochoa SEV-2012-0234. 

The results were derived with CosmoPy
(\href{http://www.ifa.hawaii.edu/cosmopy}{www.ifa.hawaii.edu/cosmopy})
and Healpix with Healpy (\href{http://healpix.jpl.nasa.gov}{healpix.jpl.nasa.gov},
\href{http://code.google.com/p/healpy}{code.google.com/p/healpy}).

Funding for SDSS-III has been provided by the Alfred P. Sloan Foundation, the Participating Institutions, the National Science Foundation, and the U.S. Department of Energy Office of Science. The SDSS-III web site is http://www.sdss3.org/.
SDSS-III is managed by the Astrophysical Research Consortium for the Participating Institutions of the SDSS-III Collaboration including the University of Arizona, the Brazilian Participation Group, Brookhaven National Laboratory, Carnegie Mellon University, University of Florida, the French Participation Group, the German Participation Group, Harvard University, the Instituto de Astrofisica de Canarias, the Michigan State/Notre Dame/JINA Participation Group, Johns Hopkins University, Lawrence Berkeley National Laboratory, Max Planck Institute for Astrophysics, Max Planck Institute for Extraterrestrial Physics, New Mexico State University, New York University, Ohio State University, Pennsylvania State University, University of Portsmouth, Princeton University, the Spanish Participation Group, University of Tokyo, University of Utah, Vanderbilt University, University of Virginia, University of Washington, and Yale University.

}

\bibliographystyle{mn2e_adslinks}
\bibliography{isw,boss,planck,powerspec}

\begin{thebibliography}{57}
\expandafter\ifx\csname natexlab\endcsname\relax\def\natexlab#1{#1}\fi

\bibitem[{{Afshordi}(2004)}]{Afshordi04}
\href{http://adsabs.harvard.edu/abs/2004PhRvD..70h3536A}{{Afshordi} N., 2004,
  \prd, 70, 083536}

\bibitem[{{Afshordi} \& {Tolley}(2008)}]{Afshordi08}
\href{http://adsabs.harvard.edu/abs/2008PhRvD..78l3507A}{{Afshordi} N.,
  {Tolley} A.~J., 2008, \prd, 78, 123507}

\bibitem[{{Aiola}, {Kosowsky} \& {Wang}(2015){Aiola}, {Kosowsky}, \&
  {Wang}}]{Aiola15}
\href{http://adsabs.harvard.edu/abs/2015PhRvD..91d3510A}{{Aiola} S., {Kosowsky}
  A., {Wang} B., 2015, \prd, 91, 043510}

\bibitem[{{Alam} {et~al}\mbox{.}(2015){Alam}, {Albareti}, {Allende Prieto},
  {Anders}, {Anderson}, {Anderton}, {Andrews}, {Armengaud}, {Aubourg},
  {Bailey}, \& et~al.}]{DR12}
\href{http://adsabs.harvard.edu/abs/2015ApJS..219...12A}{{Alam} S. {et~al.},
  2015, \apjs, 219, 12}

\bibitem[{{Anderson} {et~al}\mbox{.}(2014){Anderson}, {Aubourg}, {Bailey},
  {Beutler}, {Bhardwaj}, {Blanton}, {Bolton}, {Brinkmann}, {Brownstein},
  {Burden}, {Chuang}, {Cuesta}, {Dawson}, {Eisenstein}, {Escoffier}, {Gunn},
  {Guo}, {Ho}, {Honscheid}, {Howlett}, {Kirkby}, {Lupton}, {Manera},
  {Maraston}, {McBride}, {Mena}, {Montesano}, {Nichol}, {Nuza}, {Olmstead},
  {Padmanabhan}, {Palanque-Delabrouille}, {Parejko}, {Percival}, {Petitjean},
  {Prada}, {Price-Whelan}, {Reid}, {Roe}, {Ross}, {Ross}, {Sabiu}, {Saito},
  {Samushia}, {S{\'a}nchez}, {Schlegel}, {Schneider}, {Scoccola}, {Seo},
  {Skibba}, {Strauss}, {Swanson}, {Thomas}, {Tinker}, {Tojeiro}, {Maga{\~n}a},
  {Verde}, {Wake}, {Weaver}, {Weinberg}, {White}, {Xu}, {Y{\`e}che}, {Zehavi},
  \& {Zhao}}]{Anderson14}
\href{http://adsabs.harvard.edu/abs/2014MNRAS.441...24A}{{Anderson} L.
  {et~al.}, 2014, \mnras, 441, 24}

\bibitem[{{Bolton} {et~al}\mbox{.}(2012){Bolton}, {Schlegel}, {Aubourg},
  {Bailey}, {Bhardwaj}, {Brownstein}, {Burles}, {Chen}, {Dawson}, {Eisenstein},
  {Gunn}, {Knapp}, {Loomis}, {Lupton}, {Maraston}, {Muna}, {Myers}, {Olmstead},
  {Padmanabhan}, {P{\^a}ris}, {Percival}, {Petitjean}, {Rockosi}, {Ross},
  {Schneider}, {Shu}, {Strauss}, {Thomas}, {Tremonti}, {Wake}, {Weaver}, \&
  {Wood-Vasey}}]{Bolton12}
\href{http://adsabs.harvard.edu/abs/2012AJ....144..144B}{{Bolton} A.~S.
  {et~al.}, 2012, AJ, 144, 144}

\bibitem[{{Cabr{\'e}} \& {Gazta{\~n}aga}(2009)}]{Cabre09}
\href{http://adsabs.harvard.edu/abs/2009MNRAS.396.1119C}{{Cabr{\'e}} A.,
  {Gazta{\~n}aga} E., 2009, \mnras, 396, 1119}

\bibitem[{{Cai} {et~al}\mbox{.}(2010){Cai}, {Cole}, {Jenkins}, \&
  {Frenk}}]{Cai10}
\href{http://adsabs.harvard.edu/abs/2010MNRAS.407..201C}{{Cai} Y.-C., {Cole}
  S., {Jenkins} A., {Frenk} C.~S., 2010, \mnras, 407, 201}

\bibitem[{{Cai} {et~al}\mbox{.}(2014){Cai}, {Neyrinck}, {Szapudi}, {Cole}, \&
  {Frenk}}]{Cai14}
\href{http://adsabs.harvard.edu/abs/2014ApJ...786..110C}{{Cai} Y.-C.,
  {Neyrinck} M.~C., {Szapudi} I., {Cole} S., {Frenk} C.~S., 2014, \apj, 786,
  110}

\bibitem[{{Chon} {et~al}\mbox{.}(2004){Chon}, {Challinor}, {Prunet}, {Hivon},
  \& {Szapudi}}]{spice}
\href{http://esoads.eso.org/abs/2004MNRAS.350..914C}{{Chon} G., {Challinor} A.,
  {Prunet} S., {Hivon} E., {Szapudi} I., 2004, \mnras, 350, 914}

\bibitem[{{Crittenden} \& {Turok}(1996)}]{Crittenden96}
\href{http://adsabs.harvard.edu/abs/1996PhRvL..76..575C}{{Crittenden} R.~G.,
  {Turok} N., 1996, Physical Review Letters, 76, 575}

\bibitem[{{Einasto} {et~al}\mbox{.}(2014){Einasto}, {Lietzen}, {Tempel},
  {Gramann}, {Liivam{\"a}gi}, \& {Einasto}}]{Einasto14}
\href{http://adsabs.harvard.edu/abs/2014A%26A...562A..87E}{{Einasto} M.,
  {Lietzen} H., {Tempel} E., {Gramann} M., {Liivam{\"a}gi} L.~J., {Einasto} J.,
  2014, \aap, 562, A87}

\bibitem[{{Flender}, {Hotchkiss} \& {Nadathur}(2013){Flender}, {Hotchkiss}, \&
  {Nadathur}}]{Flender13}
\href{http://adsabs.harvard.edu/abs/2013JCAP...02..013F}{{Flender} S.,
  {Hotchkiss} S., {Nadathur} S., 2013, \jcap, 2, 13}

\bibitem[{{Foreman-Mackey} {et~al}\mbox{.}(2013){Foreman-Mackey}, {Hogg},
  {Lang}, \& {Goodman}}]{emcee}
\href{http://adsabs.harvard.edu/abs/2013PASP..125..306F}{{Foreman-Mackey} D.,
  {Hogg} D.~W., {Lang} D., {Goodman} J., 2013, \pasp, 125, 306}

\bibitem[{{Francis} \& {Peacock}(2010)}]{Francis10}
\href{http://adsabs.harvard.edu/abs/2010MNRAS.406...14F}{{Francis} C.~L.,
  {Peacock} J.~A., 2010, \mnras, 406, 14}

\bibitem[{{Frommert} \& {En{\ss}lin}(2009)}]{Frommert09}
\href{http://adsabs.harvard.edu/abs/2009MNRAS.395.1837F}{{Frommert} M.,
  {En{\ss}lin} T.~A., 2009, \mnras, 395, 1837}

\bibitem[{{Giannantonio} {et~al}\mbox{.}(2010){Giannantonio}, {Martinelli},
  {Silvestri}, \& {Melchiorri}}]{Giannantonio10}
\href{http://esoads.eso.org/abs/2010JCAP...04..030G}{{Giannantonio} T.,
  {Martinelli} M., {Silvestri} A., {Melchiorri} A., 2010, \jcap, 4, 30}

\bibitem[{{Giannantonio} {et~al}\mbox{.}(2014){Giannantonio}, {Ross},
  {Percival}, {Crittenden}, {Bacher}, {Kilbinger}, {Nichol}, \&
  {Weller}}]{Giannantonio14}
\href{http://esoads.eso.org/abs/2014PhRvD..89b3511G}{{Giannantonio} T., {Ross}
  A.~J., {Percival} W.~J., {Crittenden} R., {Bacher} D., {Kilbinger} M.,
  {Nichol} R., {Weller} J., 2014, \prd, 89, 023511}

\bibitem[{{Giannantonio} {et~al}\mbox{.}(2008){Giannantonio}, {Scranton},
  {Crittenden}, {Nichol}, {Boughn}, {Myers}, \& {Richards}}]{Giannantonio08}
\href{http://adsabs.harvard.edu/abs/2008PhRvD..77l3520G}{{Giannantonio} T.,
  {Scranton} R., {Crittenden} R.~G., {Nichol} R.~C., {Boughn} S.~P., {Myers}
  A.~D., {Richards} G.~T., 2008, \prd, 77, 123520}

\bibitem[{{G{\'o}rski} {et~al}\mbox{.}(2005){G{\'o}rski}, {Hivon}, {Banday},
  {Wandelt}, {Hansen}, {Reinecke}, \& {Bartelmann}}]{Healpix}
\href{http://adsabs.harvard.edu/abs/2005ApJ...622..759G}{{G{\'o}rski} K.~M.,
  {Hivon} E., {Banday} A.~J., {Wandelt} B.~D., {Hansen} F.~K., {Reinecke} M.,
  {Bartelmann} M., 2005, ApJ, 622, 759}. \url{http://healpix.jpl.nasa.gov/}

\bibitem[{{Granett}, {Neyrinck} \& {Szapudi}(2008){Granett}, {Neyrinck}, \&
  {Szapudi}}]{Granett08}
\href{http://adsabs.harvard.edu/abs/2008ApJ...683L..99G}{{Granett} B.~R.,
  {Neyrinck} M.~C., {Szapudi} I., 2008, \apjl, 683, L99}

\bibitem[{{Granett}, {Neyrinck} \& {Szapudi}(2009){Granett}, {Neyrinck}, \&
  {Szapudi}}]{Granett09}
\href{http://adsabs.harvard.edu/abs/2009ApJ...701..414G}{---, 2009, \apj, 701,
  414}

\bibitem[{{Granett, B. R.} {et~al}\mbox{.}(2015){Granett, B. R.}, {Branchini,
  E.}, {Guzzo, L.}, {Abbas, U.}, {Adami, C.}, {Arnouts, S.}, {Bel, J.},
  {Bolzonella, M.}, {Bottini, D.}, {Cappi, A.}, {Coupon, J.}, {Cucciati, O.},
  {Davidzon, I.}, {De Lucia, G.}, {de la Torre, S.}, {Fritz, A.}, {Franzetti,
  P.}, {Fumana, M.}, {Garilli, B.}, {Ilbert, O.}, {Iovino, A.}, {Krywult, J.},
  {Le Brun, V.}, {Le Fèvre, O.}, {Maccagni, D.}, {Małek, K.}, {Marulli, F.},
  {McCracken, H. J.}, {Polletta, M.}, {Pollo, A.}, {Scodeggio, M.}, {Tasca, L.
  A. M.}, {Tojeiro, R.}, {Vergani, D.}, {Zanichelli, A.}, {Burden, A.}, {Di
  Porto, C.}, {Marchetti, A.}, {Marinoni, C.}, {Mellier, Y.}, {Moutard, T.},
  {Moscardini, L.}, {Nichol, R. C.}, {Peacock, J. A.}, {Percival, W. J.}, \&
  {Zamorani, G.}}]{Granett15}
{Granett, B. R.} {et~al.}, 2015, \aap, 583, A61.
  \url{http://dx.doi.org/10.1051/0004-6361/201526330}

\bibitem[{{Hern{\'a}ndez-Monteagudo}
  {et~al}\mbox{.}(2014){Hern{\'a}ndez-Monteagudo}, {Ross}, {Cuesta},
  {G{\'e}nova-Santos}, {Xia}, {Prada}, {Rossi}, {Neyrinck}, {Viel},
  {Rubi{\~n}o-Martin}, {Sc{\'o}ccola}, {Zhao}, {Schneider}, {Brownstein},
  {Thomas}, \& {Brinkmann}}]{HM14}
\href{http://adsabs.harvard.edu/abs/2014MNRAS.438.1724H}{{Hern{\'a}ndez-Monteagudo}
  C. {et~al.}, 2014, \mnras, 438, 1724}

\bibitem[{{Hern{\'a}ndez-Monteagudo} \& {Smith}(2013)}]{HM13}
\href{http://adsabs.harvard.edu/abs/2013MNRAS.435.1094H}{{Hern{\'a}ndez-Monteagudo}
  C., {Smith} R.~E., 2013, \mnras, 435, 1094}

\bibitem[{{Ho} {et~al}\mbox{.}(2012){Ho}, {Cuesta}, {Seo}, {de Putter}, {Ross},
  {White}, {Padmanabhan}, {Saito}, {Schlegel}, {Schlafly}, {Seljak},
  {Hern{\'a}ndez-Monteagudo}, {S{\'a}nchez}, {Percival}, {Blanton}, {Skibba},
  {Schneider}, {Reid}, {Mena}, {Viel}, {Eisenstein}, {Prada}, {Weaver},
  {Bahcall}, {Bizyaev}, {Brewinton}, {Brinkman}, {Nicolaci da Costa}, {Gott},
  {Malanushenko}, {Malanushenko}, {Nichol}, {Oravetz}, {Pan},
  {Palanque-Delabrouille}, {Ross}, {Simmons}, {de Simoni}, {Snedden}, \&
  {Yeche}}]{Ho12}
\href{http://adsabs.harvard.edu/abs/2012ApJ...761...14H}{{Ho} S. {et~al.},
  2012, \apj, 761, 14}

\bibitem[{{Ho} {et~al}\mbox{.}(2008){Ho}, {Hirata}, {Padmanabhan}, {Seljak}, \&
  {Bahcall}}]{Ho08}
\href{http://adsabs.harvard.edu/abs/2008PhRvD..78d3519H}{{Ho} S., {Hirata} C.,
  {Padmanabhan} N., {Seljak} U., {Bahcall} N., 2008, \prd, 78, 043519}

\bibitem[{{Hotchkiss} {et~al}\mbox{.}(2015){Hotchkiss}, {Nadathur},
  {Gottl{\"o}ber}, {Iliev}, {Knebe}, {Watson}, \& {Yepes}}]{Hotchkiss15}
\href{http://adsabs.harvard.edu/abs/2015MNRAS.446.1321H}{{Hotchkiss} S.,
  {Nadathur} S., {Gottl{\"o}ber} S., {Iliev} I.~T., {Knebe} A., {Watson} W.~A.,
  {Yepes} G., 2015, \mnras, 446, 1321}

\bibitem[{{Ili{\'c}} {et~al}\mbox{.}(2011){Ili{\'c}}, {Douspis}, {Langer},
  {P{\'e}nin}, \& {Lagache}}]{Ilic11}
\href{http://adsabs.harvard.edu/abs/2011MNRAS.416.2688I}{{Ili{\'c}} S.,
  {Douspis} M., {Langer} M., {P{\'e}nin} A., {Lagache} G., 2011, \mnras, 416,
  2688}

\bibitem[{{Ili{\'c}}, {Langer} \& {Douspis}(2013){Ili{\'c}}, {Langer}, \&
  {Douspis}}]{Ilic13}
\href{http://adsabs.harvard.edu/abs/2013A%26A...556A..51I}{{Ili{\'c}} S.,
  {Langer} M., {Douspis} M., 2013, \aap, 556, A51}

\bibitem[{{Kov{\'a}cs} \& {Granett}(2015)}]{Kovacs15}
\href{http://adsabs.harvard.edu/abs/2015MNRAS.452.1295K}{{Kov{\'a}cs} A.,
  {Granett} B.~R., 2015, \mnras, 452, 1295}

\bibitem[{{Manzotti} \& {Dodelson}(2014)}]{Manzotti14}
\href{http://adsabs.harvard.edu/abs/2014PhRvD..90l3009M}{{Manzotti} A.,
  {Dodelson} S., 2014, \prd, 90, 123009}

\bibitem[{{Maturi} {et~al}\mbox{.}(2007){Maturi}, {Dolag}, {Waelkens},
  {Springel}, \& {En{\ss}lin}}]{Maturi07}
\href{http://adsabs.harvard.edu/abs/2007A%26A...476...83M}{{Maturi} M., {Dolag}
  K., {Waelkens} A., {Springel} V., {En{\ss}lin} T., 2007, \aap, 476, 83}

\bibitem[{{Nadathur} \& {Hotchkiss}(2014)}]{Nadathur14}
\href{http://adsabs.harvard.edu/abs/2014MNRAS.440.1248N}{{Nadathur} S.,
  {Hotchkiss} S., 2014, \mnras, 440, 1248}

\bibitem[{{Nadathur}, {Hotchkiss} \& {Sarkar}(2012){Nadathur}, {Hotchkiss}, \&
  {Sarkar}}]{Nadathur12}
\href{http://adsabs.harvard.edu/abs/2012JCAP...06..042N}{{Nadathur} S.,
  {Hotchkiss} S., {Sarkar} S., 2012, \jcap, 6, 42}

\bibitem[{{Nadathur} {et~al}\mbox{.}(2014){Nadathur}, {Lavinto}, {Hotchkiss},
  \& {R{\"a}s{\"a}nen}}]{Nadathur14cs}
\href{http://adsabs.harvard.edu/abs/2014PhRvD..90j3510N}{{Nadathur} S.,
  {Lavinto} M., {Hotchkiss} S., {R{\"a}s{\"a}nen} S., 2014, \prd, 90, 103510}

\bibitem[{{Navarro}, {Frenk} \& {White}(1996){Navarro}, {Frenk}, \&
  {White}}]{Navarro96}
\href{http://adsabs.harvard.edu/abs/1996ApJ...462..563N}{{Navarro} J.~F.,
  {Frenk} C.~S., {White} S.~D.~M., 1996, \apj, 462, 563}

\bibitem[{{Neyrinck}(2008)}]{Neyrinck08}
\href{http://adsabs.harvard.edu/abs/2008MNRAS.386.2101N}{{Neyrinck} M.~C.,
  2008, \mnras, 386, 2101}

\bibitem[{{Neyrinck}, {Gnedin} \& {Hamilton}(2005){Neyrinck}, {Gnedin}, \&
  {Hamilton}}]{Neyrinck05}
\href{http://adsabs.harvard.edu/abs/2005MNRAS.356.1222N}{{Neyrinck} M.~C.,
  {Gnedin} N.~Y., {Hamilton} A.~J.~S., 2005, \mnras, 356, 1222}

\bibitem[{{Nishizawa}(2014)}]{Nishizawa14}
\href{http://adsabs.harvard.edu/abs/2014PTEP.2014fB110N}{{Nishizawa} A.~J.,
  2014, Progress of Theoretical and Experimental Physics, 2014, 060000}

\bibitem[{{Peacock} \& {Dodds}(1994)}]{PeacockDodds94}
\href{http://adsabs.harvard.edu/abs/1994MNRAS.267.1020P}{{Peacock} J.~A.,
  {Dodds} S.~J., 1994, \mnras, 267, 1020}

\bibitem[{{Peiris}(2014)}]{Peiris2014}
{Peiris} H.~V., 2014, in IAU Symposium, Vol. 306, IAU Symposium, pp. 124--130

\bibitem[{{Planck Collaboration} {et~al}\mbox{.}(2015{\natexlab{a}}){Planck
  Collaboration}, {Adam}, {Ade}, {Aghanim}, {Akrami}, {Alves}, {Arnaud},
  {Arroja}, {Aumont}, {Baccigalupi}, \& et~al.}]{Planck}
\href{http://adsabs.harvard.edu/abs/2015arXiv150201582P}{{Planck Collaboration}
  {et~al.}, 2015{\natexlab{a}}, ArXiv e-prints}

\bibitem[{{Planck Collaboration} {et~al}\mbox{.}(2015{\natexlab{b}}){Planck
  Collaboration}, {Adam}, {Ade}, {Aghanim}, {Arnaud}, {Ashdown}, {Aumont},
  {Baccigalupi}, {Banday}, {Barreiro}, \& et~al.}]{PlanckMaps}
\href{http://adsabs.harvard.edu/abs/2015arXiv150205956P}{---,
  2015{\natexlab{b}}, ArXiv e-prints}

\bibitem[{{Planck Collaboration} {et~al}\mbox{.}(2015{\natexlab{c}}){Planck
  Collaboration}, {Ade}, {Aghanim}, {Arnaud}, {Ashdown}, {Aumont},
  {Baccigalupi}, {Banday}, {Barreiro}, {Bartolo}, \& et~al.}]{PlanckISW}
\href{http://adsabs.harvard.edu/abs/2015arXiv150201595P}{---,
  2015{\natexlab{c}}, ArXiv e-prints}

\bibitem[{{Raccanelli} {et~al}\mbox{.}(2015){Raccanelli}, {Kovetz}, {Dai}, \&
  {Kamionkowski}}]{Raccanelli15}
\href{http://adsabs.harvard.edu/abs/2015arXiv150203107R}{{Raccanelli} A.,
  {Kovetz} E., {Dai} L., {Kamionkowski} M., 2015, ArXiv e-prints}

\bibitem[{{Rassat}, {Starck} \& {Dup{\'e}}(2013){Rassat}, {Starck}, \&
  {Dup{\'e}}}]{Rassat13}
\href{http://adsabs.harvard.edu/abs/2013A%26A...557A..32R}{{Rassat} A.,
  {Starck} J.-L., {Dup{\'e}} F.-X., 2013, \aap, 557, A32}

\bibitem[{{Rees} \& {Sciama}(1968)}]{reessciama}
\href{http://adsabs.harvard.edu/abs/1968Natur.217..511R}{{Rees} M.~J., {Sciama}
  D.~W., 1968, \nat, 217, 511}

\bibitem[{{Ross} {et~al}\mbox{.}(2012){Ross}, {Percival}, {S{\'a}nchez},
  {Samushia}, {Ho}, {Kazin}, {Manera}, {Reid}, {White}, {Tojeiro}, {McBride},
  {Xu}, {Wake}, {Strauss}, {Montesano}, {Swanson}, {Bailey}, {Bolton}, {Dorta},
  {Eisenstein}, {Guo}, {Hamilton}, {Nichol}, {Padmanabhan}, {Prada},
  {Schlegel}, {Maga{\~n}a}, {Zehavi}, {Blanton}, {Bizyaev}, {Brewington},
  {Cuesta}, {Malanushenko}, {Malanushenko}, {Oravetz}, {Parejko}, {Pan},
  {Schneider}, {Shelden}, {Simmons}, {Snedden}, \& {Zhao}}]{Ross12}
\href{http://adsabs.harvard.edu/abs/2012MNRAS.424..564R}{{Ross} A.~J. {et~al.},
  2012, \mnras, 424, 564}

\bibitem[{{Sachs} \& {Wolfe}(1967)}]{sachswolfe}
\href{http://adsabs.harvard.edu/abs/1967ApJ...147...73S}{{Sachs} R.~K., {Wolfe}
  A.~M., 1967, \apj, 147, 73}

\bibitem[{{Sawangwit} {et~al}\mbox{.}(2010){Sawangwit}, {Shanks}, {Cannon},
  {Croom}, {Ross}, \& {Wake}}]{Sawangwit10}
\href{http://esoads.eso.org/abs/2010MNRAS.402.2228S}{{Sawangwit} U., {Shanks}
  T., {Cannon} R.~D., {Croom} S.~M., {Ross} N.~P., {Wake} D.~A., 2010, \mnras,
  402, 2228}

\bibitem[{{Sutter} {et~al}\mbox{.}(2014){Sutter}, {Pisani}, {Wandelt}, \&
  {Weinberg}}]{Sutter14}
\href{http://adsabs.harvard.edu/abs/2014MNRAS.443.2983S}{{Sutter} P.~M.,
  {Pisani} A., {Wandelt} B.~D., {Weinberg} D.~H., 2014, \mnras, 443, 2983}

\bibitem[{{Swanson} {et~al}\mbox{.}(2008){Swanson}, {Tegmark}, {Hamilton}, \&
  {Hill}}]{Mangle}
\href{http://adsabs.harvard.edu/abs/2008MNRAS.387.1391S}{{Swanson} M.~E.~C.,
  {Tegmark} M., {Hamilton} A.~J.~S., {Hill} J.~C., 2008, MNRAS, 387, 1391}

\bibitem[{{Szapudi} {et~al}\mbox{.}(2015){Szapudi}, {Kov{\'a}cs}, {Granett},
  {Frei}, {Silk}, {Burgett}, {Cole}, {Draper}, {Farrow}, {Kaiser}, {Magnier},
  {Metcalfe}, {Morgan}, {Price}, {Tonry}, \& {Wainscoat}}]{Szapudi15}
\href{http://adsabs.harvard.edu/abs/2015MNRAS.450..288S}{{Szapudi} I. {et~al.},
  2015, \mnras, 450, 288}

\bibitem[{{Takahashi} {et~al}\mbox{.}(2012){Takahashi}, {Sato}, {Nishimichi},
  {Taruya}, \& {Oguri}}]{Takahashi12}
\href{http://adsabs.harvard.edu/abs/2012ApJ...761..152T}{{Takahashi} R., {Sato}
  M., {Nishimichi} T., {Taruya} A., {Oguri} M., 2012, \apj, 761, 152}

\bibitem[{{Umetsu} {et~al}\mbox{.}(2015){Umetsu}, {Zitrin}, {Gruen}, {Merten},
  {Donahue}, \& {Postman}}]{Umetsu15}
\href{http://adsabs.harvard.edu/abs/2015arXiv150704385U}{{Umetsu} K., {Zitrin}
  A., {Gruen} D., {Merten} J., {Donahue} M., {Postman} M., 2015, ArXiv
  e-prints}

\bibitem[{{Zibin}(2014)}]{Zibin14}
\href{http://adsabs.harvard.edu/abs/2014arXiv1408.4442Z}{{Zibin} J.~P., 2014,
  ArXiv e-prints}

\end{thebibliography}

\label{lastpage}

\end{document}